# Electronic properties of ordered and disordered linear clusters of atoms and molecules[†]

**S N Behera**[a], **S Gayen**[b], G **V Ravi Prasad**[c] and **S M Bose**[b,*]

[a]Institute of Material Science, Bhubaneswar-751013, Orissa, India.
[b]Department of Physics, Drexel University, 3147 Chestnut Street, Philadelphia, PA 19104, USA.
[c]Institute of Physics, Sachivalaya Marg, Bhubaneswar 751005, Orissa, India.

## Abstract

The electronic properties of one-dimensional clusters of $N$ atoms or molecules have been studied. The model used is similar to the Kronig-Penney model with the potential offered by each ion being approximated by an attractive $\delta$-function. The energy eigenvalues, the eigenstates and the density of states are calculated exactly for a linear cluster of $N$ atoms or molecules. The dependence of these quantities on the various parameters of the problem show interesting behavior. Effects of random distribution of the positions of the atoms and random distribution of the strengths of the potential have also been studied. The results obtained in this paper can have direct applications for linear chain of atoms produced on metal surfaces or artificially created chain of atoms by using scanning tunneling microscope or in studying molecular conduction of electrons across one-dimensional barriers.

# 1. Introduction

With the advances in nanoscience and nanotechnology, the study of the electronic properties of a nanoparticle or a cluster has attained much importance in recent times. These studies are interesting not only because they show the intrinsic properties of these systems but also because they exhibit the localized properties of an extended solid. For example, it is important to know a cluster of how many atoms or molecules will start forming an energy band and begin to behave like a bulk material. The loss of atomic like behavior and evolution of the band structure with the number of atoms or molecules have been studied by various authors by various methods [1-7]. Most of these techniques are computation intensive and hence can deal satisfactorily with only a relatively small number of constituent atoms or molecules. It would be interesting both from scientific and pedagogical value to use a model in which the problem can be solved analytically in closed form. Such a method would allow us to obtain insight into the properties of a cluster which can then be extended to get solution for a cluster of arbitrary size.

A lot of insight has been gained into the electronic structure of solids by looking at one-dimensional models which are exactly solvable. One such model which can be found in every text book on solid state physics is the Kronig-Penney model [8]. In this model the atoms, in the one-dimensional crystalline solid, are represented by the square well potentials, for which the Schrödinger equation can be solved exactly. In the limit this problem reduces to the periodic delta-function potential which admits analytic solution [9]. The great advantage of this reduction is that the band structure which is so crucial to the understanding of the electronic properties of solids can be calculated analytically and its dependence on the potential parameters can be evaluated explicitly. The problem has been dealt with using both the repulsive as well as the attractive delta-function potentials [10]. However it is well known that the attractive delta-functions will have only one band of bound states whereas that of the repulsive delta-functions will have an infinite number of positive energy bands [12]. One can visualize that the attractive and repulsive periodic δ-function potentials will represent, respectively, the tight binding and the nearly free electron pictures of the band structure.

In spite of considerable development in the understanding of the electronic properties of solids, certain fundamental questions still remain unanswered. One such question relates to the emergence of bulk properties of solids, e.g., the electrical conductivity, the refractive index, the magnetic susceptibility, the elastic constants etc.; while single atoms are devoid of such properties. The question then is at the aggregation of how many number of atoms do such properties arise. The answer to this question is expected to emerge from the study of atomic clusters and nanomaterials. Since the one-dimensional periodic δ-function models have been useful in providing clues to our understanding of the electronic properties of solids; it is only natural to expect that even the answer to this question can be sought for within this model. The advantage of this model is that any number of δ-function potentials can be solved exactly – one and two attractive δ-functions mimicking as models for a single atom and a diatomic molecule, respectively, has been demonstrated [11] successfully. So it would be interesting to extend the model and apply it to a cluster of N atoms. Starting from this point of view, in this paper we calculate the eigenstates, eigenenergies and the densities of states for electrons of a cluster comprised of a chain of an arbitrary number of atoms or molecules, where the interaction of the electrons with the ions is idealized by attractive δ-function potentials and where the atoms or molecules in the cluster are either free or contained by two infinitely high potential barriers at the two ends.

The main motivation for our calculation is to simulate the formation of clusters of atoms or molecules and study their electronic behavior as the number of atoms or molecules in the cluster increases. Although such a one-dimensional calculation may not be quite realistic, but it will provide a simple way of determining theoretically, for example, how many atoms are necessary to form a band of electronic states and how this band structure will evolve with the change in the number of atoms or molecules in the cluster and how they will depend on parameters such as the distance between the consecutive constituent atoms, on the location of the boundaries of the cluster and the strengths of the atomic potentials.

In recent times the molecular conduction of electrons across one-dimensional barriers has become important because of its potential application in nanotechnology [13]. We expect that for such calculations, the theory presented in this paper will have important relevance. Furthermore, using the manipulative power of a scanning tunneling microscope, one-dimensional chains of atoms and molecules are now being prepared [14]. The calculations presented in the current paper should be useful in providing qualitative features of the band structure, density of states, effective mass, etc. of these artificially produced one-dimensional metals and semiconductors.

The organization of the rest of the paper is as follows: In section 2 the model and its solution for a linear chain of N-atomic cluster represented by periodically arranged n-attractive δ-function potentials will be presented. One can conceive of two kinds of clusters: (i) free atomic clusters and (ii) bound clusters. The free clusters are defined as those on which no extra boundary condition is imposed on the two end atoms of the cluster. On the other hand a bound cluster is one, at the two ends of which there are infinite potential walls which confine the cluster within a square well potential. The former case of the free atomic cluster has been investigated earlier [15] in some detail. In this paper we will concentrate on the bound clusters of atoms and molecules. Only some representative results will be presented for the free case, which will enable the comparison of the similarities and differences between the free and bound clusters. It will be shown that band like features start emerging even with as small a number as 8 atoms for a cluster. These results for the clusters of N atoms are presented in Section 2.1. The bound clusters are expected to behave like nanomaterials [16]. They have the versatility of depicting not only the single tight binding band corresponding to the bound state of the attractive δ-function potential but also the positive energy continuum states (Section 2.2) which give rise to nearly free electron like bands. Besides, the emergence of surface states (also called end states) can be clearly seen for the bound clusters. Section 3 is devoted to the discussion of molecular clusters. This can be achieved by replacing the periodic variation of the positions of *N* attractive δ-functions of the same strength by *N/2* diatomic molecules the periodic arrangement of which resembles the dimerization of the lattice. The other possibility is to consider a cluster of hetero-diatomic molecule consisting of two different types of atoms, which can be simulated by choosing attractive δ-functions with two different strengths. The results for both these cases will be presented again for the tight binding band of bound states as well as the positive energy continuum states. Section 4 is devoted to a study of disordered clusters. Again there can be two different types of disorder in a cluster: (i) where there is a random variation of the strengths of the attractive δ-function potential keeping the spacing fixed and (ii) where there is a random variation in the spacing between consecutive δ-functions, while their strength remains the same. The first problem is analogous to the Anderson localization [17] in a cluster, and the second corresponds to an amorphous cluster [18]. For both kinds of disorder results will be presented both for the tight binding band of bound states as well as the nearly free electron bands corresponding to the positive energy continuum. It will be shown that the two kinds of disorder have totally different effects on the tight binding and the nearly free electron bands. Section 5 is devoted to a brief summary of the results and their discussion.

**2. Model for a cluster of atoms**

In this section we present a model to study the band structure of electrons confined in a one-dimensional cluster of $N$ atoms and present its exact solution. To start with, the theory is worked out for the bound states (negative energy solutions) of electrons in an $N$ particle cluster represented by $N$ periodic δ-function potentials of a constant negative strength with infinite walls at the two ends. The energy eigenvalues, eigenfunctions and the density of electronic states are calculated as a function of the number of atoms $N$ as well as the other parameters of the problem such as the strength of the potential, the separation between consecutive atoms and the distance of the boundaries from the end atoms. Next the energy eigenvalues and density of states corresponding to the positive energy solutions for the electrons of a cluster of $N$ atoms are calculated. But first the general method for the exact solution of the problem is presented; which consist of solving the Schrödinger equation with appropriate boundary conditions.

**2.1 Bound states (negative energy) of electrons in a linear cluster of N atoms**

The atomic cluster or the nanomaterial under consideration consisting of $N$ atoms is defined by $N$ attractive δ-function potentials of a given strength *($a_0$)* and separated by a distance $b$ from each other. It is also assumed that the electrons are subjected at the two ends to infinite potential barriers and these barriers are at a distance $a_1$ and $a_2$ from the first and last atoms (δ-functions) of the cluster, respectively. As an example, this configuration is depicted in Figure 1 for a cluster of 4 atoms. The solution of the problem consists of solving the Schrödinger equation for negative energy corresponding to the bound states of the electron.

We first write down the solution of the Schrödinger equation for one and two δ-function potentials and then generalize these solutions to the case of $N$ δ-functions by using the method of induction. For the case of one δ-function, the Schrödinger equation for an electron of mass $m$ has the well known form

$$-\frac{\hbar^2}{2m}\frac{d^2}{dx^2}\psi(x)+V(x)\psi(x)=E\psi(x) \tag{1}$$

where $V(x)$ represents the δ-function potential placed at $x=a_1$ and $\psi$ and $E$ are the usual eigenfunction and eigenenergy of the electron. If we define $V(x)\equiv-\frac{\hbar^2}{2m}a_0\delta(x)$ and $E\equiv\frac{\hbar^2}{2m}\varepsilon$, where $\varepsilon$ must be negative, i.e., $\varepsilon=-k^2$, since we are looking for a negative energy solution, Eq. (1) takes a somewhat simpler form

$$\frac{d^2\psi(x)}{dx^2}-k^2\psi(x)=-a_0\delta(x-b)\psi(x) \tag{2}$$

The boundary conditions required to be satisfied by the wave function are that it must vanish at the two ends of the cluster and be continuous at the location of the δ-functions, but its slope must have a discontinuity determined by the strength of the δ-function, $a_0$. So the boundary conditions for a single δ-function at $a_1$ are

$$\begin{aligned}&\psi(x=0)=0\\&\psi(x=a_1)=continuous\\&\frac{d\psi}{dx}\bigg|_{a_1+\delta}-\frac{d\psi}{dx}\bigg|_{a_1-\delta}=-a_0\psi(a_1)\\&\psi(x=a_1+a_2)=0\end{aligned} \tag{3}$$

where $a_1 \pm \delta$ are the distances just after and just before $a_1$ by infinitesimal amounts. Calling the regions to the left and right of the δ-function as regions 1 and 2, respectively, we can write the solution of Eq. (2) in the two regions as

$$\psi(x) = A_1 e^{kx} + B_1 e^{-kx} \quad \text{for} \quad 0 < x < a_1$$
$$\psi(x) = A_2 e^{kx} + B_2 e^{-kx} \quad \text{for} \quad a_1 < x < a_1 + a_2 \tag{4}$$

Using the boundary conditions given by Eqs. (3) on the wave functions of Eq. (4), we obtain the following relationships between the coefficients $A_i$'s and $B_i$'s in the two regions:

$$A_1 + B_1 = 0$$
$$A_2 = (1 - \frac{a_0}{2k}) A_1 - \frac{a_0}{2k} Y^{-2} B_1$$
$$B_2 = \frac{a_0}{2k} Y^2 A_1 + (1 + \frac{a_0}{2k}) B_1$$
$$A_2 YZ + B_2 Y^{-1} Z^{-1} = 0$$

(5)

where the quantities Y and Z are defined as $Y \equiv e^{ka_1}$ and $Z \equiv e^{ka_2}$.

If we now proceed to do the same calculation for the case of two δ-function potentials separated by a distance $b$, we will obtain the following Schrödinger equation using the definitions described above

$$\frac{d^2 \psi(x)}{dx^2} - k^2 \psi(x) = -a_0 [\delta(x - a_1) + \delta(x - (a_1 + b))] \psi(x)$$

(6)

Here the region of interest can be divided into three regions: region 1 to the left of the first δ-function ($0 < x < a_1$), region 2 between the two δ-functions ($a_1 < x < a_1 + b$) and region 3 to the right of the second δ-function ($a_1+b < x < a_1+b+a_2$). The boundary conditions in this case will get modified to

$$\begin{aligned}
&\psi(x=0)=0 \\
&\psi(x=a_1)=continuous \\
&\frac{d\psi}{dx}\Big|_{a_1+\delta} - \frac{d\psi}{dx}\Big|_{a_1-\delta} = -a_0\psi(a_1) \\
&\psi(x=a_1+b)=continuous \\
&\frac{d\psi}{dx}\Big|_{a_1+b+\delta} - \frac{d\psi}{dx}\Big|_{a_1+b-\delta} = -a_0\psi(a_1+b) \\
&\psi(x=a_1+b+a_2)=0
\end{aligned} \quad (7)$$

The solutions of the Schrödinger equation in the three regions can be expressed as linear combinations of the exponential function $e^{\pm kx}$ with coefficients $A_i$'s and $B_i$'s ($i=1,2,3$) and can be written as

$$\begin{aligned}
\psi(x) &= A_1 e^{kx} + B_1 e^{-kx} \quad \text{for} \quad 0 < x < a_1 \\
\psi(x) &= A_2 e^{kx} + B_2 e^{-kx} \quad \text{for} \quad a_1 < x < a_1+b \\
\psi(x) &= A_3 e^{kx} + B_3 e^{-kx} \quad \text{for} \quad a_1+b < x < a_1+b+a_2
\end{aligned} \quad (8)$$

Applying the boundary conditions given in Eqs. (7) to the wave functions given in Eqs. (8) we obtain the following relationships among the coefficients $A_i$'s and $B_i$'s

$$\begin{aligned}
&A_1 + B_1 = 0 \\
&A_2 = (1 - \frac{a_0}{2k})A_1 - \frac{a_0}{2k} Y^{-2} B_1 \\
&B_2 = \frac{a_0}{2k} Y^2 A_1 + (1 + \frac{a_0}{2k}) B_1 \\
&A_3 = (1 - \frac{a_0}{2k})A_2 - \frac{a_0}{2k} Y^{-2} X^{-2} B_2 \\
&B_3 = \frac{a_0}{2k} Y^2 X^2 A_2 + (1 + \frac{a_0}{2k}) B_2 \\
&A_3 XYZ + B_3 X^{-1} Y^{-1} Z^{-1} = 0
\end{aligned} \quad (9)$$

where $X \equiv e^{kb}$. The first and the last equations are the consequences of the fact that the wave function should vanish at the two boundaries. Note that $A_2$ and $B_2$ are exactly the same as those for one δ-function case and the coefficients in the third region $A_3$ and $B_3$ have similar forms as those of $A_2$ and $B_2$ except for the introduction of the multiplicative factor $X^{\pm 2}$ in appropriate places. If we extend the calculation to a cluster of three atoms represented by three δ-functions, the space would be divided into four regions. The coefficients in the second and third regions will be exactly the same as those given in Eqs. (9) and the coefficients in the fourth region, by induction, will be given by

$$A_4 = (1 - \frac{a_0}{2k}) A_3 - \frac{a_0}{2k} Y^{-2} X^{-4} B_3$$
$$B_4 = \frac{a_0}{2k} Y^2 X^4 A_1 + (1 + \frac{a_0}{2k}) B_3$$
(10)

and the last equation of Eqs. (9) will be modified to

$$A_4 X^2 YZ + B_4 X^{-2} Y^{-1} Z^{-1} = 0 .$$

Note that in this case $A_4$ and $B_4$ have the same forms as those of $A_3$ and $B_3$ except for another set of multiplicative factor of $X^{\pm 2}$ in appropriate places. We can then generalize this calculation to the case of any number of δ-functions by induction, and the coefficients in any of the *(i+1)*-th region can be expressed in terms of the coefficients in the *i*-th region by an appropriate multiplication of $X^{\pm 2}$.

For a cluster of *N* atoms represented by *N* δ-function potentials, we will thus have the following relationships among the coefficients of the wave functions.

$$A_1 + B_1 = 0$$
$$\vdots$$
$$A_{N+1} = (1 - \frac{a_0}{2k}) A_N - \frac{a_0}{2k} Y^{-2} X^{-2(N-1)} B_N$$
$$B_{N+1} = \frac{a_0}{2k} Y^2 X^{2(N-1)} A_N + (1 + \frac{a_0}{2k}) B_N$$
$$A_{N+1} Y X^{(N-1)} Z + B_{N+1} Y^{-1} X^{-(N-1)} Z^{-1} = 0$$
(11)

We reiterate that the first and the last equations in Eqs. (11) are obtained from the fact that the wave function must go to zero at the two ends of the confined $N$ atom cluster. The energy eigenvalues ($k^2$) for this problem can be obtained from Eq. (11). The basic idea is to express $A_{N+1}$ and $B_{N+1}$ in terms $A_1$ or $B_1$ by utilizing the relationships given in Eqs. (11) and then use these expressions in the last equation to obtain an equation for $A_1$ or $B_1$. The energy eigenvalues are found by setting the coefficient of $A_1$ or $B_1$ to be zero. Since all the coefficients can be written in terms of $A_1$ or $B_1$, the only unknown coefficient is then $A_1$ or $B_1$. To find this coefficient we utilize the normalization condition of the wave function, i.e.

$$\int_0^{a_1 + (N-1)b + a_2} \psi^* \psi \, dx = 1 \tag{12}$$

This calculation is straightforward and can be done analytically. Once all the coefficients are known, the wave functions can then be calculated in all the segments of the cluster.

We will now present the results of our calculation. The results will obviously depend on the parameters $N$, $a_1$, $a_2$, $b$ and $a_0$, i.e, the number of atoms, the distance of the first and last δ-function from the two boundaries, the separation between consecutive δ-functions and the strength of the attractive δ-function, respectively. In Figure 2 the energy eigenvalues for several clusters with the number of atoms *N=4, 8, 16, 32* and *64* are shown as horizontal lines for the choice of the parameters $a_1=a_2=b=4.0$, and $a_0= 2.0$. As can be seen, for *N=4* there are four energy eigenvalues which are well separated and spread over the energy range between –0.527 to –0.467 corresponding to a spread in energy of around 0.06 in appropriate units. However for *N=8, 16, 32* and *64* a distinct banding effect is seen to emerge. The spread in energy increases only slightly to about 0.074 as one goes to the larger clusters while the energy eigen values at the middle being uniform, and that at the top and bottom bunch together. Thus by *N=64* the band seems to be quite well formed.

The band width and the number of states depend also on the parameters *N, $a_1$, $a_2$, $a_0$* and *b*. Further analysis shows that this dependence on *b* is significant as can be seen in Figure 3a, where we have plotted the energy eigenvalues as a function of *1/b* for *N=8*. Although we would expect eight energy eigenvalues for eight δ-function potentials, we see that the actual number is dependent on *b*. For very large values of *b*, there is only one eight fold degenerate eigenstate because an electron effectively sees each δ-function potential separately, with no overlap of the wave function. As *b* decreases the degeneracy is lifted and the number of energy states becomes *8*, with the separation between consecutive levels increasing rapidly. However, for much smaller values of *b(<<1)* the eigenstates start to disappear into the positive energy continuum and eventually just one eigenstate will survive, because of the collapse of all the δ-functions to just one δ-function with 8 times the strength of the individual potential. This will be the general behavior of the eigenenergies as a function of *1/b* for all clusters with *N* atoms.

This behavior is analogous to that of a free cluster of *N* atoms [15], i.e., one without boundaries. The reason for this can be attributed to the fact that with decreasing separation between consecutive δ-function potentials *b*, the effective distance of the boundaries from their nearest atoms increases rapidly, there by making the presence of the boundaries redundant. However, the situation will not be the same if simultaneously not only *b* but $a_1$ and $a_2$ are also decreased so that the *N*-atom cluster is always contained. The variation of energy eigenvalues for the case *$a_1$= $a_2$=b*, as a function of *1/b* is shown in Figure 3b. As expected, for sufficiently small values of *b*, i.e. *1/b* around *20*, we find that all the bound energy eigenstates disappear into the continuum. This is a consequence of the uncertainty principle; as the eigenfunction gets confined in smaller and smaller regions of length, the kinetic energy of the electron increases; which overtakes the bound state energy at a certain critical length pushing the bound state into the continuum.

To show the effect of the location of the boundaries, i.e., infinitely high potential barriers at the ends of the cluster, on the energy eigenvalues and eigenstates, we have plotted in Figure 3c the energy eigenvalues for $N=8$ as a function of $1/a$ ($a=a_1=a_2$) while keeping the separation between consecutive δ-functions (b=4) as well as the strength of the potential ($a_0=2$) constants. Here again we notice that the number of allowed states depends on the magnitude of $a$. For values of $a$ greater than $b/2$ (=2), we get eight states which is equal to the number of atoms $N$ in the cluster, as in the case of a free cluster. But when $a$ becomes much smaller than $b/2$, the number of bound states decreases to six as two of the eigenstates shift to positive energies. This transition from $N$ to $(N-2)$ states is interesting. As $a$ becomes smaller only two of the states get significantly affected because at the two boundaries the wave functions must vanish, which results in the squeezing of the wave functions in the regions $0 < x < a_1$ and $a_1+(N-1)b < x < a_1+(N-1)b + a_2$, which in turn increases the energy of two of the states. These two states are the surface states (end states) of the linear cluster of $N$-atoms and can be clearly seen Figure 3c around $a \leq b/2$. For even smaller values of $a$, these two surface states come together becoming almost degenerate before moving to positive energies for $a< 0.59$. Thus the surface states can either be seen or not seen depending on the distance of the boundaries from the atoms at the two ends of the cluster. This conclusion is true for any value $N$.

Using Eqs. (11) and (12), one can also calculate the coefficients of the wave function in each region of this one-dimensional chain of atoms for given values of $N$, $a_0$, $a_1$, $a_2$, and $b$. In Figure 4a, the four wave functions are plotted for four possible energy eigenvalues ($k$) for a four atom cluster ($N=4$) with $a_0=2$, $a_1=a_2=b=4$. It is interesting to note that the wave functions for the lowest energy state and the second excited states have odd reflection symmetry whereas those of the first and third excited states have even reflection symmetry about the y-axis drawn at the centre of the cluster. For any even $N$ such symmetries will alternate with energy eigenvalues, the lowest eigenstate always having odd symmetry. In Figure 4b we have plotted the five wave functions for $N=5$. Clearly there are three even and two odd symmetry states, the lowest energy state being even and the other states alternating between odd and even.

Another quantity of importance in determining the electronic structure of materials is the density of states (DOS). The DOS enters into the calculation of many physical properties of solids and also it can be inferred from experimental measurements of some properties of solids. Therefore, it is interesting to calculate the density of states of the electrons for our system of a one-dimensional cluster of $N$ atoms.

Even though there are indications of the emergence of a band from the calculation of the energy eigenvalues for the N-atom linear cluster as shown in Figure 2, a proper signature of the formation of the band can only be obtained from the DOS. With this in mind we have carried out such a calculation by counting the number of states per unit energy (and per unit length) at each energy value and the result of our calculation for a sixty four atom cluster is shown in Figure 5. As can be easily seen, the behavior of the density of states for the 64 atom cluster already shows similarities to that of an infinitely long linear chain of atoms, with the emergence of the van Hove singularities at the two edges of the band and smooth variation in the middle of the band. The characteristics of the singularities are typical of a one-dimensional system. However, one can notice a distinct asymmetry between the singularities at the two edges, the singularity at the lower energy end being stronger than that at the upper energy end. This we believe is a typical characteristic of a cluster of a finite number of atoms in a linear chain as compared to a bulk solid, i.e., a chain with infinite number of atoms or a chain with periodic boundary conditions. Even though the van Hove singularities shown in Figure 5a are for the 64-atoms cluster, the signature of the singularities start appearing at much smaller cluster size as can be seen from Figure 5b, where we have plotted the DOS for a 16-atoms cluster. Therefore it will not be out of context to say that a cluster of size N=16 already starts exhibiting bulk properties. This essentially answers the question raised in the Introduction.

**2.2 Scattering states (positive energy) of electrons in a linear chain of N atoms**

It is interesting to ask what the scattering or continuum states of such a confined *N*-atomic cluster are. In the case of a free *N*-atomic cluster since there is no confining potential the positive energy solutions are just free electron states forming the continuum [19]. The presence of the attractive δ-functions can at best produce phase shifts and result in resonances. But the case of a confined cluster is different. In the absence of the attractive δ-function potentials the problem reduces to that of an electron in an infinite square well potential, whose eigenvalues are proportional to $n^2$ (*n* being an integer). So there is infinite number of energy eigen values or positive energy bound states. The question then is what happens to these bound states in the presence of *N*-attractive δ-function potentials of *N*-atoms of the cluster? To find the answer, in this subsection, we calculate the energy eigenvalues and eigenstates of scattering state (positive energy) solutions of the electrons confined in an infinite potential barrier in a region containing *N* atoms represented by *N* attractive equispaced δ-function potentials. The theory for this case is almost exactly the same as that discussed in the preceding subsection, except that in this case the energy ε will be positive, i.e., in Eqs. (2) and (6) $k^2$ will be replaced by $-k^2$. Hence in all equations *k* will be replaced *ik*. For example, for a two atom cluster the solution of Schrödinger equation given by Eq. (8) will be replaced by

$$\begin{aligned}\psi(x) &= A_1 e^{ikx} + B_1 e^{-ikx} \quad \text{for} \quad 0 < x < a_1 \\ \psi(x) &= A_2 e^{ikx} + B_2 e^{-ikx} \quad \text{for} \quad a_1 < x < a_1 + b \\ \psi(x) &= A_3 e^{ikx} + B_3 e^{-ikx} \quad \text{for} \quad a_1 + b < x < a_1 + b + a_2\end{aligned} \tag{13}$$

For the general case of *N* atoms, applying the appropriate boundary conditions at the locations of the δ-functions and at the boundaries of the cluster we would get the following equations for the coefficients $A_i$ and $B_i$:

$$A_1 + B_1 = 0$$
$$\vdots$$
$$A_{N+1} = \left(1 - \frac{a_0}{2ik}\right) A_N - \frac{a_0}{2ik} Y^{-2} X^{-2(N-1)} B_N$$
$$B_{N+1} = \frac{a_0}{2ik} Y^2 X^{2(N-1)} A_N + \left(1 + \frac{a_0}{2ik}\right) B_N$$
$$A_{N+1} Y X^{(N-1)} Z + B_{N+1} Y^{-1} X^{-(N-1)} Z^{-1} = 0$$
(14)

where $X \equiv e^{ikb}$, $Y \equiv e^{ika_1}$, and $Z \equiv e^{ika_2}$. These equations are very similar to Eqs. (11) where *k* has been replaced by *ik*. The normalization condition given in Eq. (12) will be satisfied in this case as well. Using Eqs. (14) and Eq. (12), we can calculate the energy eigenvalues and eigenstates exactly the same way as discussed in the preceding subsection.

The results of this calculation are presented in Figures 6 to 8. In Figure 6 we have plotted the energy eigen values for the clusters of sizes *N=4, 8, 16, 32*, and *64* for the case of $a_0=2$, and $a_1=a_2=b=4$. As expected for the scattering states, we obtain infinitely many positive energy solutions. However, we notice that these energy values start forming bands separated by band gaps which are clearly visible for clusters starting with *N=8* and the band structure is quite prominent by the time the cluster size reaches *N=64*.

Some features of these energy bands depicted in Figure 6 are worth noting. Unlike the nearly free electron bands, in this case the lowest band does not start with zero energy, but starts at a finite energy. This is the quantum mechanical zero point energy and it has its origin in the fact that the infinite square well potential does not admit a solution corresponding to the integer *n=0*; the lowest energy eigenvalue being the one corresponding to *n=1*. So the band structure starts with a band gap. The band width goes on increasing as one goes to the upper bands, while the band gaps more or less remain constant. There is a clear bunching of states at the band edges, the bunching being more prominent at the lower band edges. The average energy of each of these bands are around *0.5, 1.7, 3.5, 6 …* corresponding to *n=1 ,2 ,3, 4 …*, respectively. However, for the infinite square well potential these energies are expected to be at *1, 4, 9, 16 …* etc. There is a distinct lowering of the energies of the bands; because of the presence of the attractive δ-functions within the square well. As a consequence there is an effective lowering of the bottom of the well to a value below the energy zero. The grouping of states for the formation of the bands arises due to the lifting of the degeneracy of the states due to the overlap of the wave functions in the regions in between the δ-function potentials. Thus the origin of the bands for the positive energy solutions is the same as that of the bound state. These bands can be thought of as analogous to the nearly free electron picture while the band of bound states is related to the tight binding picture. Some of the bands in Figure 6 seem to contain *(N+1)* states rather *N* states as would be expected. The presence of *(N+1)* states in a band depends on the relative values *a* and *b* as will be discussed below. The band gaps depend only on the strength of the attractive δ-function. We find that increasing $a_0$ increases the band gap, the upper edge of a band is pushed down whereas the lower edge does not move appreciably, thus increasing the band gap.

In Figure 7 we have plotted the density of states for several lower bands for the cluster with *N=64*. As expected each band shows the van Hove singularities at each edge, the characteristic of a one-dimensional system. The asymmetry in the singularities at the two edges persists.

The band structure will obviously depend on the values of the parameters such as $N$, $a_0$, $b$, $a_1$ and $a_2$. In Figures 8a and 8b we plot the energy distribution of the lower bands as a function of $1/a$ (with $a=a_1=a_2$) and $b$, respectively. Figure 8a shows the evolution of the band structure as the boundaries move from far off to closer and closer to the two end point δ-functions, while the separation between the consecutive δ-function is kept fixed at b=4. When $a$ is large, say, $a=10$ or slightly less (i.e. $1/a = 0.1$ or slightly more) the cluster is expected to behave like a free cluster so far as its bound states are concerned (see discussion in sec. 2.1), because the wave functions in the regions of the boundaries decay exponentially. But the positive energy solutions being oscillatory in nature will persist no matter how far the boundaries are. However, for $a=10$ and $b=4$ for the *8*-atom cluster the width of the infinite square well potential $L$ will be $L=2a+(N-1)b = 48$. Since the energy eigenvalues are proportional to $1/L^2$, the separation between consecutive energy values will be smaller and smaller with increasing $L$. As a result for small $1/a$ the band structure will not be apparent as can be seen from Figure 8a. In bringing the boundaries closer to the edges of the cluster (i) first of all the separation between consecutive energy values will increase and (ii) the vanishing of the wave function at the boundary will result in the splitting off of the top most two energy eigen values to higher energies; resulting in surface states. Such emergence of surface states from the lower band and their merging with the bottom of the upper band can be seen in Figure 8a. Consequently depending on the choice of the value of *1/a* a band can have either *7* or *9* or *8* states as can be seen from Figure 8a. This may even result in bands having alternately *7* and *9* states for the *8*-atom cluster or in general *(N-1)* and *(N+1)* states in an *N*-atom cluster. This explains the observation of *(N+1)* states per band in Figure 6. Also it is clear that by the time *1/a>1*, a clear band structure emerges. This discussion demonstrates the invalidity of the usual notion about the origin of the band structure that it is due to the perfect periodicity of the lattice arising from the periodic boundary condition.

Variation of energy eigenvalues with *b* shows similar behavior as shown in Figure 8b. The separation of bands becomes blurred for small values of *b* and there is migration of states from one band to the other with variation of *b*. For small values of *b* there is large overlap of wave functions, resulting in large splitting of energy eigenvalues, which blurs the formation of bands. The bands emerge for *b>2*. The band width decreases as b increases. This figure again shows that as a consequence of the presence of the boundaries, two surface states originate at the top of each band and merge with the bottom of the next upper band. Depending on the value of *b* the bands can have either (*N-1*) or (*N+1*) states for an *N*-atom cluster. When the band has (*N-1*) states there will be two surface states in the mid gap region. These are the general features of the nearly free electron bands of an *N*-atom cluster.

## 3. Molecular clusters

In this section we generalize our calculation to a cluster of molecules in the linear chain where each molecule can have more than one atom. First we consider the case of diatomic molecules and then indicate how to generalize this to the case of n atoms per molecule. For the case of *N* periodically placed diatomic molecules in the linear chain, we note that each molecule is represented by two attractive δ-function potentials of strengths $-a_{01}$ and $-a_{02}$, respectively. The separation between two atoms within the same molecule is taken to be $b_1$ and that between successive molecules is taken to be $b_2$, the latter separation being similar to the lattice constant. The distances of the first atom of the first molecule and last atom of the last molecule from the two boundaries are, as before, $a_1$ and $a_2$. As usual the boundaries are represented by two infinite walls, which contain the cluster. Here again we use the usual boundary conditions that at the location of each δ-function, the wave function is continuous but its slope is discontinuous and the magnitude of the discontinuity is determined by the strength of the corresponding δ-function, and that the wave function goes to zero at the two boundaries of the cluster. For a cluster of *N* diatomic molecules, the total length of the cluster will be divided into (*2N+1*) regions. Using these boundary conditions, we get the following equations relating the coefficients of the wave functions in the different regions for the negative energy of the electrons:

$$A_1 + B_1 = 0$$
$$\vdots$$

$$A_{2N} = (1 - \frac{a_0}{2k}) A_{2N-1} - \frac{a_0}{2k} Y^{-2} (X_1 X_2)^{-2(N-1)} B_{2N-1}$$

$$B_{2N} = \frac{a_0}{2k} Y^2 (X_1 X_2)^{2(N-1)} A_{2N-1} + (1 + \frac{a_0}{2k}) B_{2N-1} \qquad (15)$$

$$A_{2N+1} = (1 - \frac{a_{02}}{2k}) A_{2N} - \frac{a_{02}}{2k} Y^{-2} X_1^{-2} (X_1 X_2)^{-2(N-1)} B_{2N}$$

$$B_{2N+1} = \frac{a_{02}}{2k} Y^2 X_1^2 (X_1 X_2)^{2(N-1)} A_{2N} + (1 + \frac{a_{02}}{2k}) B_{2N}$$

$$A_{2N+1} Y X_1^N X_2^{N-1} Z + B_{2N+1} Y^{-1} X_1^{-n} X_2^{-(N-1)} Z^{-1} = 0$$

where $X_1 \equiv e^{kb_1}$, $X_2 = e^{kb_2}$, and $Y$ and $Z$ retain their meaning as have been defined immediately below Eqs. (5). As in the previous two cases, Eq. (15) along with Eq. (12) can be used to obtain the energy eigenvalues and eigenfunction for the cluster of $N$ diatomic molecules. The result of our calculation for the bound state (negative energy) energies of electrons in a thirty two diatomic molecular cluster is shown in Figure 9a. We note that for $b_1 = b_2 = 4$ and $a_{01} = a_{02} = 2$, the system corresponds to a cluster of sixty four atoms and in this case we obtain just one band (the leftmost band), which as expected is identical with the one shown in Figure 2. The cases where $b_1$ and $b_2$ have different values, and the strengths $a_{01} = a_{02}$ are the same, results in the $N=32$ homo molecular diatomic clusters and this results in two energy bands. This separation of a band into two is very similar to what one obtains when one calculates the band structure of one-dimensional dimerized infinite crystal of diatomic molecules [20]. Because of the larger value of the lattice constant $b_2$ the overlap of the wave function of consecutive molecules will be smaller which will result in the reduction in the splitting of the consecutive energy levels of the molecule. Consequently, the band width is expected to be reduced. The results shown in Figure 9a depict the reduction in the band width and increase in the band gap as the separation between successive molecules $b_2$ increases. In Figure 9b we have plotted the variation of the energy bands of a cluster of thirty two diatomic molecules with variation of the strength $a_{02}$ from $a_{01} = 2.0$ while the separations between consecutive atoms as well as that between the molecules is taken to be the same, i.e., $b_1 = b_2 = 4.0$. Here again we note that when $a_{01}$ and $a_{02}$ have the same value as expected we get the single band of the sixty four atom cluster (the leftmost band, please note the different scales in Figures 9a & 9b). Variation of $a_{02}$, however, splits the band into two, since in this case the system behaves like a cluster of heterogeneous diatomic molecules even though $b_1 = b_2$. As can be seen from this figure the amount of splitting depends on the relative values of $a_{01}$ and $a_{02}$. The splitting between the two bands is much larger for a hetero molecular cluster as compared to the case of a homo molecular cluster. Similar results are obtained when $N$ is varied.

To examine the nature of the wave functions of the electrons in a diatomic linear cluster we have plotted in Figure 9c, the wave functions of a 4-homo molecular diatomic cluster for eight values of $k$. The behavior of these wave functions is similar to those of atomic clusters. Although not individually identified in the figure, we find that the lowest state has odd symmetry, and the symmetry of the higher states alternate between even and odd, very much like atomic case.

We now consider the positive energy solutions of the diatomic clusters. The equations for the coefficients of the wave functions of the electrons with positive energy eigenvalues can be obtained from Eq. (15) by replacing $k$ by $ik$, exactly as in the atomic case. Using such an equation we have computed the energy bands for the positive energy solutions in a cluster of thirty two molecules and the results are shown in Figure10. Here again we notice that when $b_1 = b_2 = 4.0$ and $a_{02} = a_{01} = 2.0$, we are actually studying a cluster of sixty four atoms and the energy bands are exactly the same as those obtained in subsection 2.2. However, as the separation between the molecules in the cluster, $b_2$, is increased keeping the other parameters fixed, each of the original bands break up into two and new bands and gaps are formed. Variation of relative values of $a_{01}$ and $a_{02}$ will similarly split the energy bands and produce more bands and gaps. This is expected as introduction of different kinds of atoms are expected to produce more bands and gaps even in a bulk solid.

This calculation can be generalized to the case of a cluster containing $N$ molecules, each consisting of $n$ ($>2$) atoms. The wave functions will satisfy the usual boundary conditions at the locations of the δ-functions and at the two boundaries of the cluster. The wave functions in different regions of the cluster can easily be obtained by generalizing Eqs. (15). The bands obtained in this case will obviously depend on $N$ as well as $n$. For example, in the case of bound state (negative energy solutions) electrons the number of bands will be *n*, separated by (*n-1*) gaps. For the positive energy (scattering state solutions) electrons, each band will break up into different sub-bands depending upon *n*.

### 4. Randomly disordered cluster

In an atomic or molecular cluster we do not expect the atoms to have a periodic distribution and even the strength of the potential offered by the atoms may not be constant and may depend on its location and environment. In this section we consider the case of disordered linear clusters arising from the (i) random distribution of the spacing between the atoms, (ii) as well as the random distribution of the strengths of the atomic potentials in an *N* particle cluster. The calculation is carried out by invoking a random number generator which would assign a value to the potential strength or to the position of an atom within a prescribed range of these values. The energy eigenvalues are then obtained by using the equations derived in the previous sections. Subsequently one can also calculate the densities of states from the energy distribution in the bands. In Figure 11 we have plotted the energy eigenvalues of the bound state electrons of a cluster of sixty four atoms where both $a_0$ and *b* have been varied randomly. The leftmost band in this figure corresponds to the regular case, i.e., equidistant distribution of atoms (*b=4*) of equal potential strength ($a_0=2$). This band corresponds to the last band in Figure 2. In the second band of this figure, the atoms are assumed to be placed at equal distances (*b*=constant) but the potential strength ($a_0$) is allowed to vary randomly between *1.9* to *2.1*. We notice that this increases the bandwidth and that the energy eigenvalues are no longer varying smoothly within the band, and band gaps seem to develop within the extended band. The third band in the figure shows that the band becomes much wider and the uneven distribution of the energy values becomes much more pronounced when $a_0$ is allowed to have values between a wider range (*1.5<$a_0$<2.5*). Formation of band gaps is quite obvious in this case. In the fourth band of this figure we plot the variation of the energy eigenvalues with random variation of the distance *b* between the atoms from *3.9* and *4.0*, keeping the strength of the atomic potential constant ($a_0=2.0$). Here also we notice that the bandwidth increases and the energy values become nonuniform but this variation is small compared to the corresponding variation in the strengths of the potential (the second band). The last band in this figure shows that the variation in energy distribution becomes larger with larger variation of b (*3.5<b<4.5*). Again this variation is not as large as the corresponding variation in $a_0$. No significant band gap appears in these two cases, only the band width increases.

Using the energy distributions of the electrons in Figure 11, we have calculated the densities of electronic states (DOS) for two of the cases of this figure. In Figure 12a, we have plotted the DOS for the case where $a_0$ is allowed to vary between *1.5* and *2.5*. Notice that the DOS of the disordered cluster no longer varies smoothly in contrast to the ordered one (*cf.* Figure 5), instead it has uneven distribution indicating that the electronic energies become non-uniformly distributed getting bunched in some energy intervals and sparse in some other regions, giving rise to band gaps (almost horizontal lines between two states in Figure 12a) not present in the ordered case. Appearance of the gaps can be clearly seen in Figure11. It should also be noted that the DOS is much reduced compared to the ordered case. It should be noted that this reduction is because of the increase in the bandwidth, although the total number of states remains constant. In Figure 12b, we have plotted DOS for the case where *b* varies between *3.5* and *4.5* with $a_0=2$. As mentioned before, although the DOS shows some non-uniformity, lowering of strength and an increase in band width, these are not as pronounced as those for the case where $a_0$ is allowed to vary. A prominent feature in both the types of disorder is the vanishing of the band edge van Hove singularities in the DOS. This is a clear pointer to the fact that the van Hove singularities are a consequence of the periodicity of the lattice atoms. Furthermore, it should be pointed out that of the two types of disorder considered here, the one with random strengths of the potential is equivalent to the problem of Anderson localization [17], whereas the one with positional disorder of the atoms is a representation of the one-dimensional amorphous solid [18]. The results of the calculation depicted for the bound state energies in Figures 11 and 12 clearly show that for the chosen values of the parameters, the effect of the disorder due to variation of atomic potentials is more dramatic compared to that of variation in the atomic positions. In the former case there is a large increase in the band width, resulting in a decrease in the average DOS and opening of band gaps. This reduction in the DOS and presence of band gaps reduce the diffusion coefficient of the electron along the lattice thus reducing the conductivity. This was the kind of localization that was predicted by the Anderson localization theories [17]. In contrast to the randomization of the strength of the potential, the randomization of the position of the atoms increases the band width only slightly as can be seen from Figures 11 and 12b. This again is in conformity with the band tailing of amorphous structures. Because of this, the average DOS is large compared to the Anderson localization problem. Thus the positional random lattice will exhibit finite conductivity corresponding to an amorphous crystal.

Similar calculations have also been carried out for the positive energy (scattering state) solutions. The variations of the energy bands with random variation of $a_0$ and $b$ for a sixty four atom cluster are shown in Figure 13. The leftmost bands correspond to the regular distribution of atoms ($b=4.0$) each with the same strength of potential ($a_0=2.0$) for a sixty four atom cluster. Note that in each of the three bands shown the energy distribution is smooth and is exactly the same as discussed in subsection 2.2 (see Figures 6 and 7). The second and the third set of bands correspond to the cases where $a_0$ is allowed to vary randomly between *1.9* and *2.1* and between *1.5* and *2.5*, respectively. The fourth and fifth bands correspond to the random variation of $b$ between *3.9* and *4.1* and between *3.5* and *4.5*, respectively. As can be seen the energy distributions become non-smooth and the band widths change with the variation of $a_0$ and $b$.

In Figures 14a and 14b we have plotted the DOS for the cases where $a_0$ is allowed to vary between *1.5* and *2.5*, and $b$ is varied between *3.5* and *4.5*, respectively. In both cases the DOS becomes less smooth compared to the ordered case (see Figure 7). However, the change both in the width of the band and the distribution of energy states is much more pronounced in the case where $b$ is allowed to vary (see Figure 14b). In this case, except for the lowest energy band all the other bands merge and form a continuum, with bunching of states in certain regions of energy and depletion of states in other regions. When $a_0$ is randomly varied, the band widths and the van Hove singularities are largely preserved for each band, but the DOS within each band gets much less smoothly varying. This latter result is in apparent contrast with the behavior of the bound state spectrum. This difference can be understood on the basis of the fact that when $b$ is randomly varied in a homo molecular diatomic cluster, for lower values of $b$ each energy value of each band moves to higher values since the size of the cluster (width of the infinite potential) decreases, whereas for larger values of $b$ each energy value of each band moves to lower energies (as explained in Sec. 2a). Thus there is widening of each band in both directions causing overlap of successive bands. For hetero molecular diatomic clusters, however, variation of $a_0$ affects only the upper edge of each band and the lower edge stays basically unaltered as stated before. Smaller $a_0$ moves it up and larger $a_0$ moves it down. Thus the random variation of $a_0$ between *1.5* and *2.5* does not substantially affect the band widths and it basically remain the same as that of a regular sample with $a_0=2.0$ as can be seen in Figures 14a and 7.

## 5. Summary and conclusions

In this paper we have studied the electronic structure of linear clusters of *N* atoms/molecules. The atomic potentials in the cluster are represented by attractive δ-function potentials of a given strength with the two boundaries of the cluster represented by infinite potential walls. The exact values of the eigenenergies and the eigenfunctions have been obtained analytically by solving the Schrödinger equation in every region of the cluster. Effects of random variation of the distance between the atoms and random variation of the strengths of atomic potentials have also been studied by using a random number generator. It is found that while the energy eigenvalues of a four atom cluster show atomic behavior, an energy band starts to form with as few as eight or sixteen atoms in the cluster. Indeed the energy band is well formed for a sixty-four atom cluster. The negative energy (tight binding) electrons in a monatomic cluster form only one band. Ordinarily the number of states in a given band is equal to the number of atoms in the cluster. However, the energy eigenvalues have interesting dependence on the separation of atoms and as well as on the distance of the boundaries of the cluster from their nearest atoms. For small values of the separation between atoms and the distance of the end atoms from the boundaries, there is migration of the atomic states to higher energies which eventually disappear into positive energy continuum. In the positive energy continuum the electronic energies do not form a quasi continuum distribution. They, in fact, form many bands each band ordinarily having N states. However, in the positive energy region also the electronic states are seen to migrate from one band to the next as the separation between atoms decreases as well as the distance of the atoms from the edges decreases. In the case of a cluster formed by diatomic molecules, two bands are formed in the negative energy range and in the positive energy range also each band splits into two as compared to the atomic case. This is the typical behavior of a linear dimerized crystal or a crystal of diatomic molecules.

The wave functions can also be calculated analytically and the results have been presented for the atomic cluster of four and five atoms. In both cases the wave functions belonging to the different eigenvalues show particular symmetries. The odd and even symmetry of the wave functions alternate, with the lowest energy eigen state having odd symmetry for $N$ even and even symmetry for $N$ odd.

The densities of states have been calculated in all cases and have been found to have the typical one-dimensional behavior exhibiting van Hove singularities at the edges of each band and a smooth variation in the middle for the ordered distribution of atoms in the cluster.

Calculation has also been carried out for the case where the atoms are not uniformly distributed in the cluster and the strengths of the atomic potential are also not constant. For the random variation of these quantities, the energy eigenvalues and the densities of sates have been calculated both for the negative and positive energy states for an $N$ atom cluster. It is found that in both cases the band width increases and the electronic states become less uniformly distributed. These variations can be related to the Anderson localization as well as the case of an amorphous crystal.

Although the model presented in this paper is that of a one-dimensional cluster, it has relevance in the study of electronic structure of clusters for several reasons. First, like the Kronig-Penney model for a bulk crystal, a simple calculation like this can become the starting point and provide guidance for the more sophisticated realistic calculations of the electronic structure of atomic or molecular clusters. For example, this calculation shows explicitly that a cluster of as few as eight or more atoms and molecules forms energy bands. Thus this calculation provides the answer to the question when a cluster of atoms or molecules starts to show bulk like behavior. Furthermore, one-dimensional clusters have been artificially created by scanning tunneling microscopy. The calculation presented in this paper will have direct applications to such clusters. This calculation will also have relevance to molecular conduction that is currently being studied in detail because of its potential applications in nanotechnology [13, 21].

**Acknowledgement**

One of us (SNB) would like to acknowledge the hospitality of the Department of Physics, Drexel University during short visits, where part of this work was carried out.

**FIGURE CAPTIONS**

1. Schematic description of a one-dimensional cluster of *4* atoms represented by *4* δ-functions of height $-a_0$ and separated by a distance *b*. Note that the two ends of the cluster present infinite potential barriers and are at distances $a_1$ and $a_2$ from the nearest atoms.

2. Negative energy eigenvalues of a cluster of *N* atoms where *N=4, 8, 16, 32* and *64* for $a_0=2.0$ and $a_1=a_2=b=4.0$. Notice that the electronic states start to form a band from *N=8*.

3a. Negative energy eigenvalues as a function of *1/b* for an *N=8* atom cluster with $a_1=a_2=4.0$ and $a_0=2.0$. Notice that for large vales of *b* there is only one degenerate state and as *b* becomes smaller, the higher energy states migrate to positive energies.

3b. Variation of energy eigenvalues as a function of *1/b* for the case $a_1=a_2=b$. For sufficiently small values of *b* all the bound state energy levels cross over to the positive energy continuum.

3c. Negative energy eigenvalues as a function of *1/a* where $a=a_1=a_2$ for an *N=8* atom cluster with *b=4* and $a_0=2$. In this case the two higher energy states ( surface states) move toward positive values as *a* decreases in value.

4a. The wave functions as a function of *x*, the distance from the left end of the cluster, for the four eigenvalues of energy for an *N=4* atom cluster for $a_1=a_2=b=4.0$ and $a_0=2.0$. Notice that the symmetry of the wave functions alternate between odd and even, the lowest one being odd.

4b. The same plot as Figure 4a except that here *N=5*. In this case three wave functions are even and two are odd, the lowest one being even.

5a. The density of the negative energy states of an N=64 atom cluster. The density of states has the usual van Hove singularities at the edges and is smoothly varying in the middle as is expected in a one-dimensional system.

5b. The density of negative energy states for an N=16 cluster. Even a cluster of 16 atoms shows signs of von Hove singularities at the edges indicating a bulk behavior.

6. Positive energy eigenvalues of a cluster of *N* atoms where *N=4, 8, 16, 32* and *64* for $a_0=2.0$ and $a_1=a_2=b=4.0$. Notice that for *N=8*, the electronic states start to form bands, and by *N=64* the bands are well formed.

7. The density of the positive energy states for several low energy bands of an $N=64$ atom cluster. The density of states has the usual van Hove singularities at the edges and smoothly varying in the middle.

8a. Positive energy eigenvalues as a function of $1/a$ where $a=a_1=a_2$ for an $N=8$ atom cluster with $b=4$ and $a_0=2$. As in the case of negative energy states, here also there is migration of eigenstates with variation of $a$.

8b. Positive energy eigenvalues as a function of $b$ for an $N=8$ atom cluster with $a_1=a_2=4.0$ and $a_0=2.0$. As $b$ varies from large values to small values, there is migration of states from lower energy bands to higher energy bands.

9a. Negative energy eigenvalues of a cluster of thirty two diatomic molecules for $a_{01}=a_{02}=2.0$ and $a_1=a_2=b_1=4.0$ and $b_2=4.0, 5.0$ and $6.0$. Notice that $b_2=4.0$ corresponds to the case of a sixty four atom cluster and we get only one band of sixty four states. For $b_2 \neq b_1$, the system behaves like a cluster of thirty two homo-diatomic molecules and we get two bands.

9b. Negative energy eigenvalues of a cluster of thirty two diatomic molecules for $a_1=a_2=b_1=b_2=4.0$ and $a_{01}=2.0$ and $a_{02}=2.0, 1.75$ and $1.5$. Notice that $a_{02}=2.0$ corresponds to the case of a sixty four atom cluster and we get only one band of sixty four states. For $a_{02} \neq a_{01}$, the system behaves like a cluster of thirty two hetero-diatomic molecules and we get two bands.

9c. The wave functions of the electrons of a four diatomic molecule. They resemble the wave functions of atomic clusters as shown in Figures 4a and 4b. Although not identified, the lowest energy state has the odd reflection symmetry.

10. Positive energy eigenvalues of a cluster of thirty two diatomic molecules for $a_{01}=a_{02}=2.0$ and $a_1=a_2=b_1=4.0$ and $b_2=4.0, 5.0, 6.0, 7.0$ and $8.0$. Notice that $b_2=4.0$ corresponds to the case of a sixty four atom cluster and we get bands corresponding to a sixty four states. For $b_2 \neq b_1$, the system behaves like a cluster of thirty two diatomic molecules and many more bands are formed..

11. Negative energy eigenvalues of a cluster of sixty four atoms where $a_0$ and $b$ are allowed to vary randomly between certain limits. The leftmost band corresponds to a regular cluster with $a_0=2.0$ and $b=4.0$. The second and third bands show that the band width and distribution of energy states vary with random variation of $a_0$. The fourth and the fifth bands show variation of band width and energy distribution with variation in $b$

12a. The density of states of the negative energy electrons for an *N=64* atom cluster with $a_1=a_2=b=4.0$ with random variation of $a_0$ between *1.5* and *2.5*.

12b. The density of states of the negative energy electrons for an *N=64* atom cluster with $a_1=a_2=4.0$ and $a_0=2.0$ with random variation of $b$ between *3.5* and *4.5*.

13 Positive energy eigenvalues of a cluster of sixty four atoms where $a_0$ and $b$ are allowed to vary randomly between certain limits. The leftmost bands corresponds to a regular cluster with $a_0=2.0$ and $b=4.0$. The second and third bands show the band width and distribution of energy states vary with random variation of $a_0$. The fourth and the fifth bands show variation of band width and energy distribution with variation in $b$.

14a. The density of states of the positive energy electrons for several low energy bands for an *N=64* atom cluster with $a_1=a_2=b=4.0$ with random variation of $a_0$ between *1.5* and *2.5*.

14b. The density of states of the positive energy electrons for several low energy bands for an *N=64* atom cluster with $a_1=a_2=4.0$ and $a_0=2.0$ with random variation of $b$ between *3.5* and *4.5*.

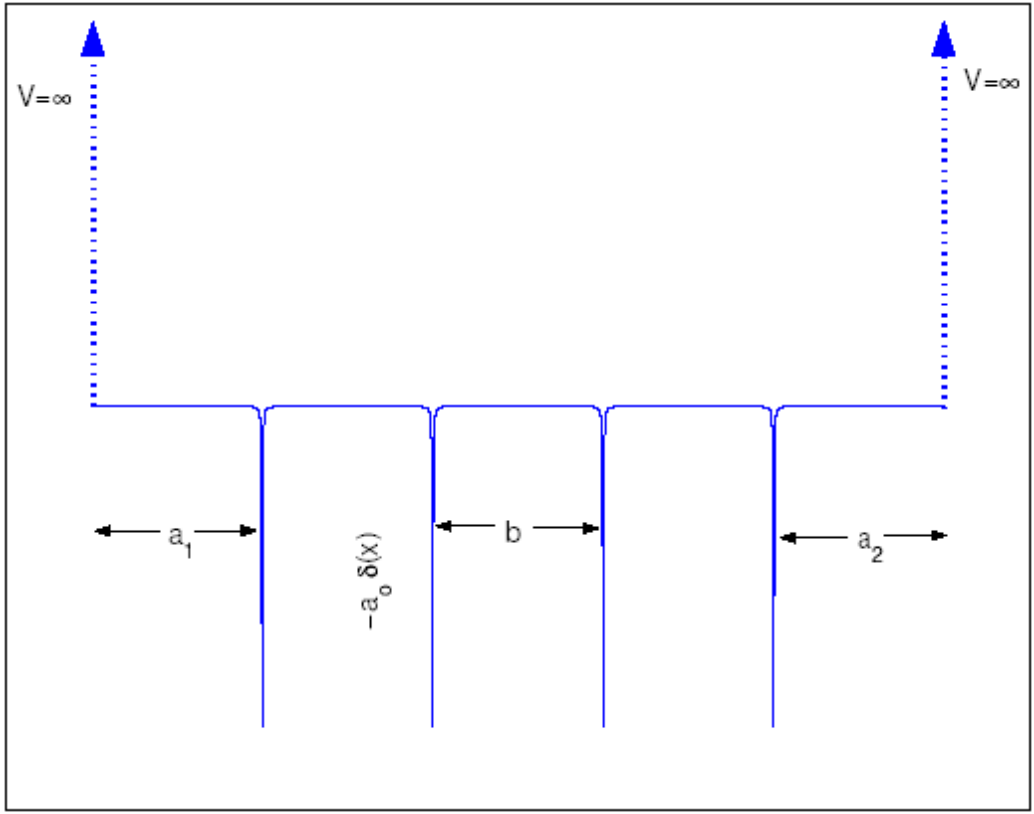

Figure 1

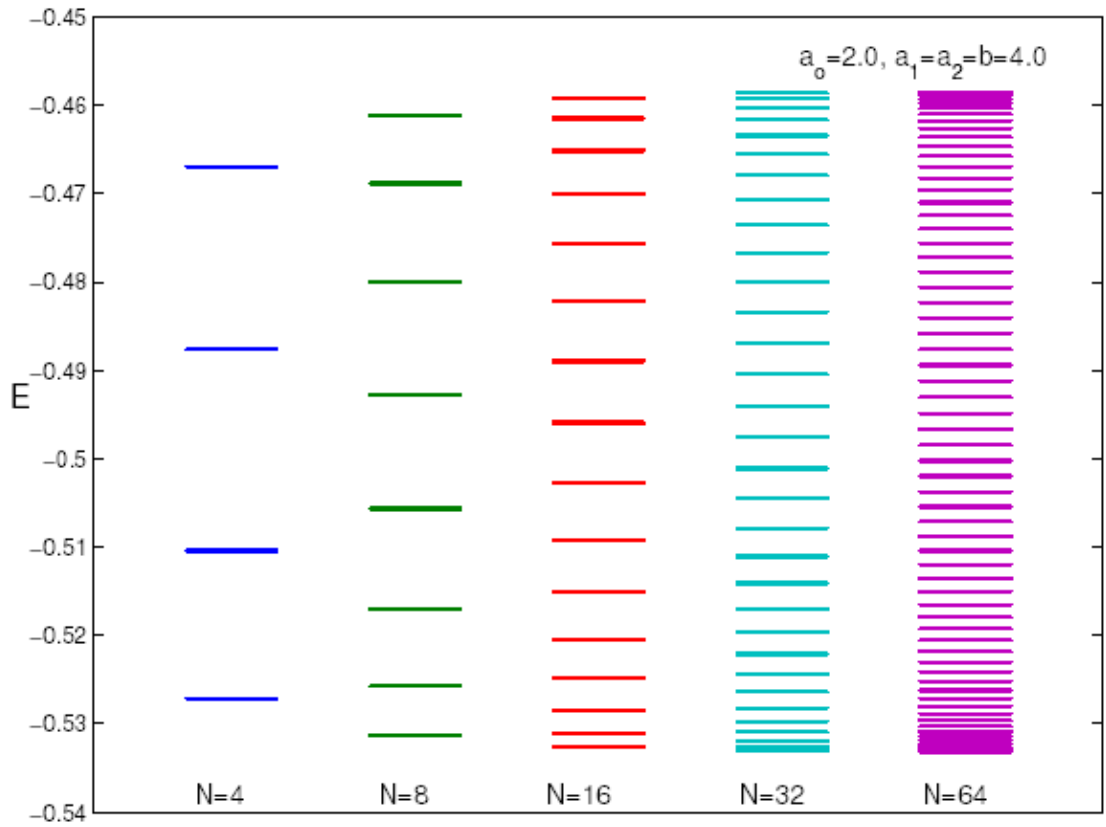

Figure 2

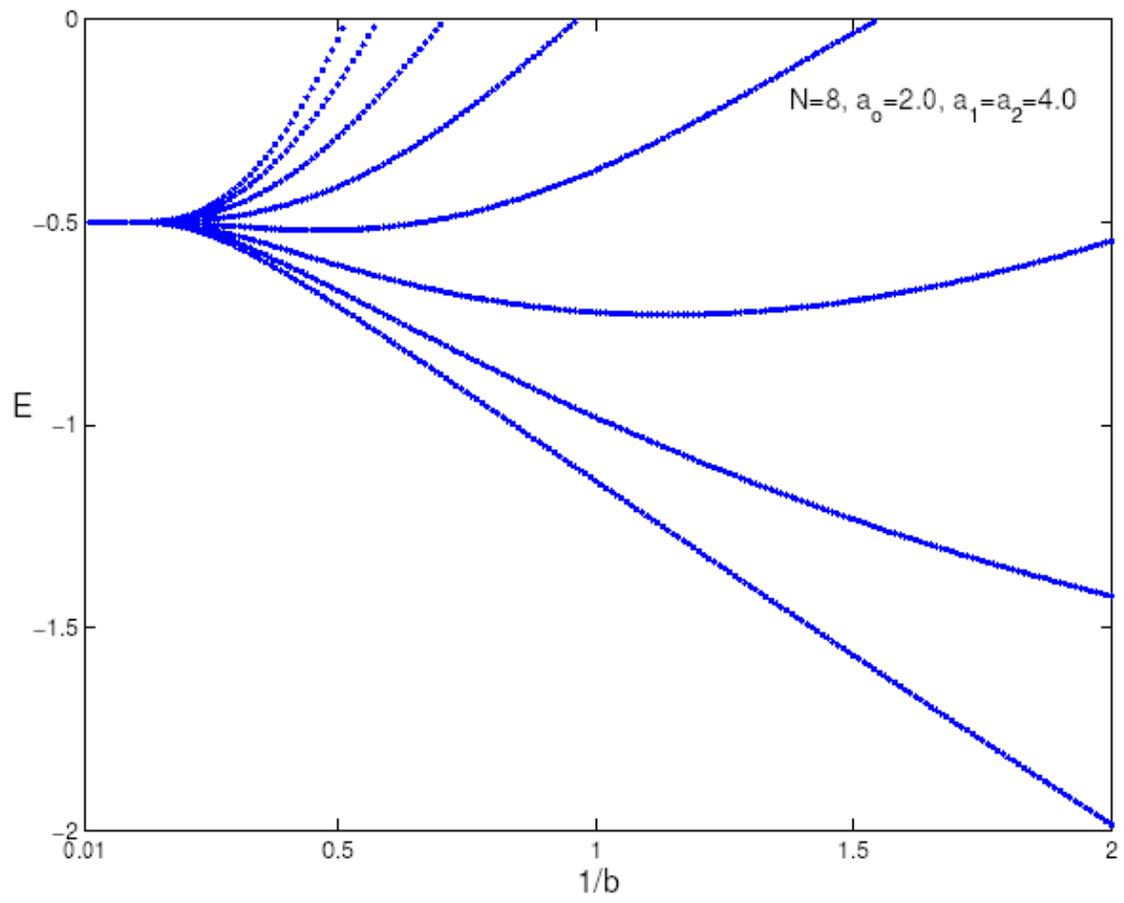

Figure 3a

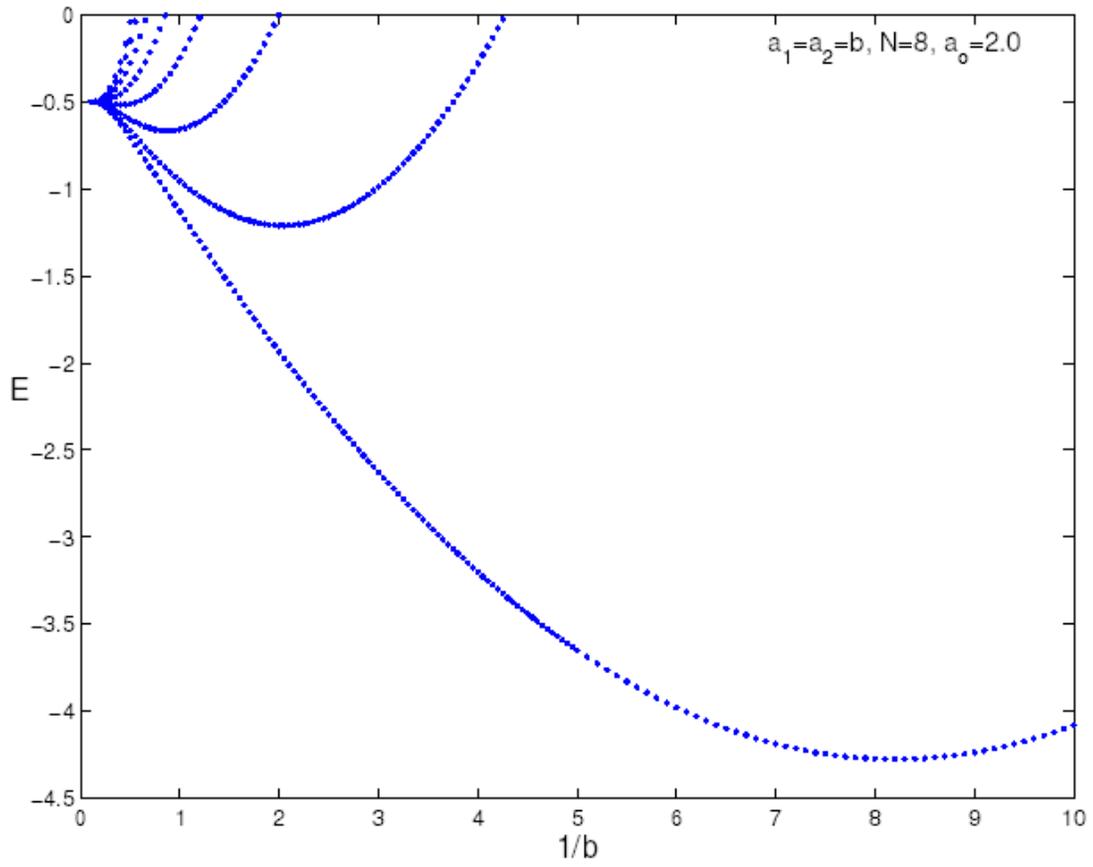

Figure 3b

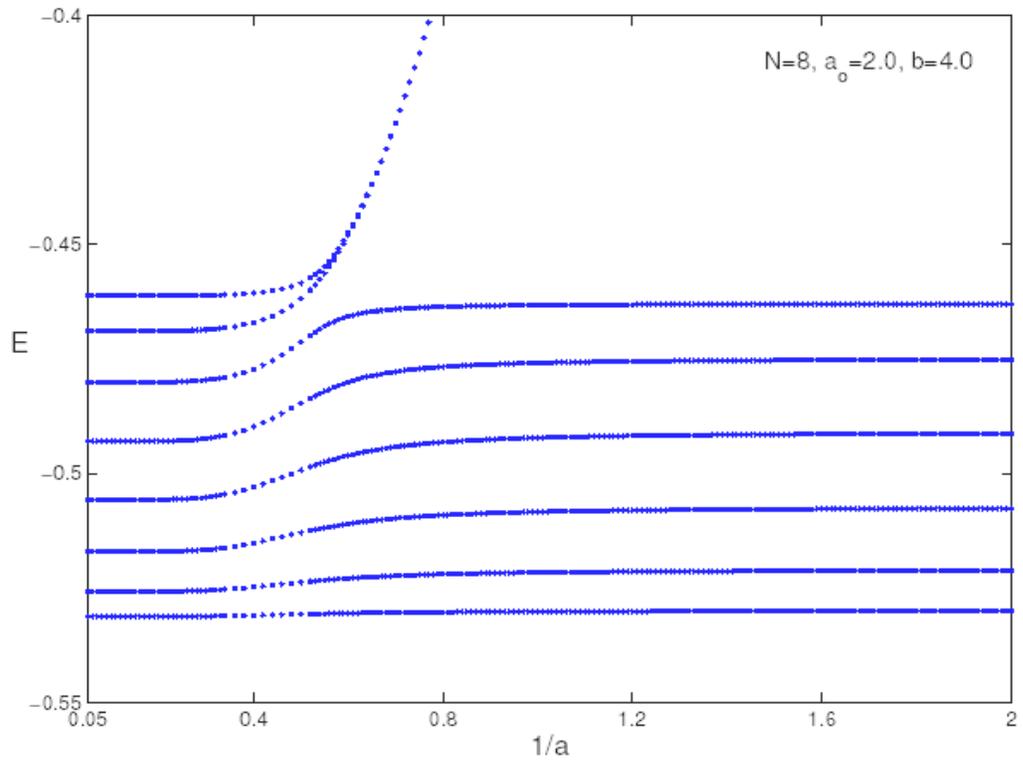

Figure 3c

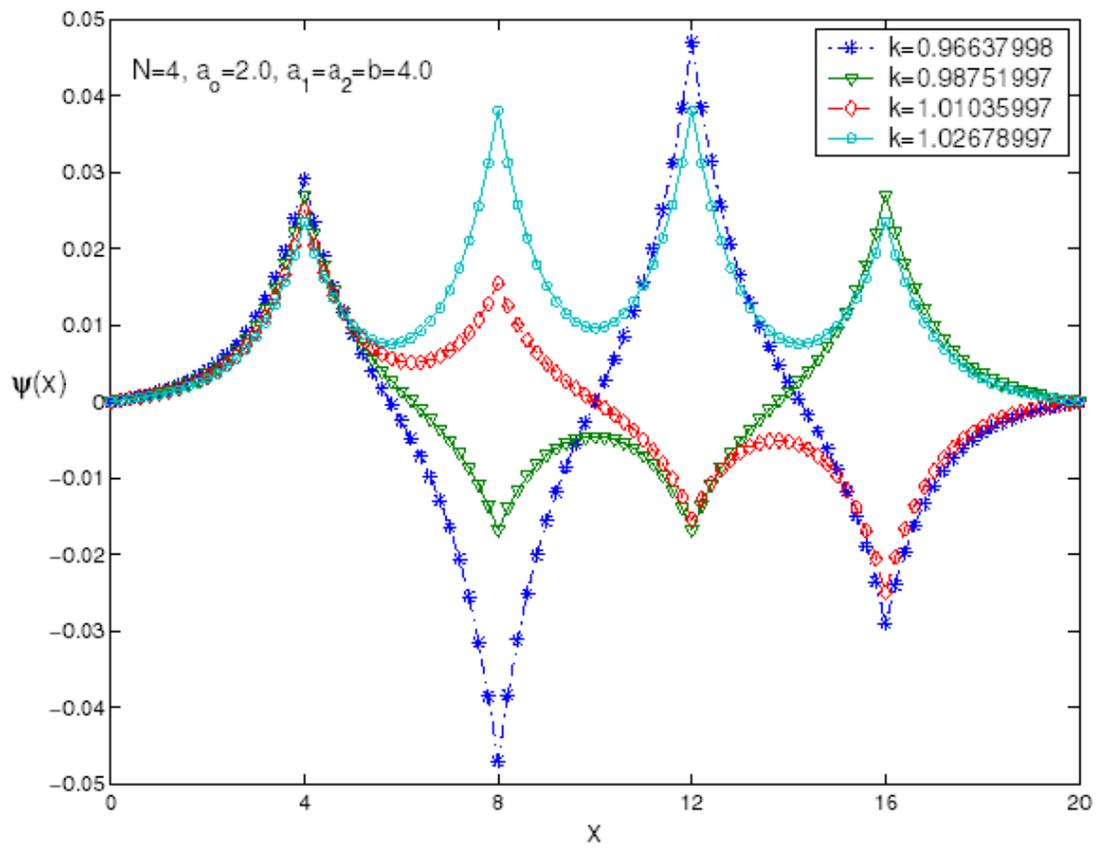

Figure 4a

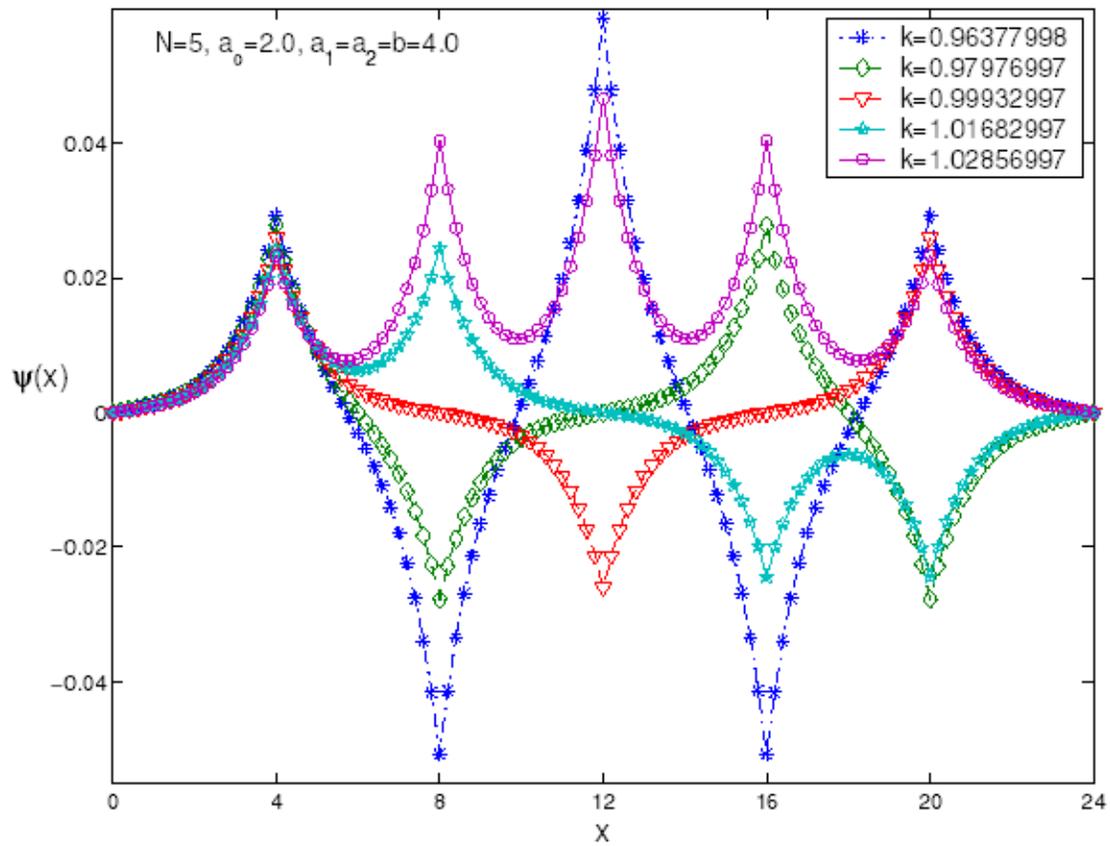

Figure 4b

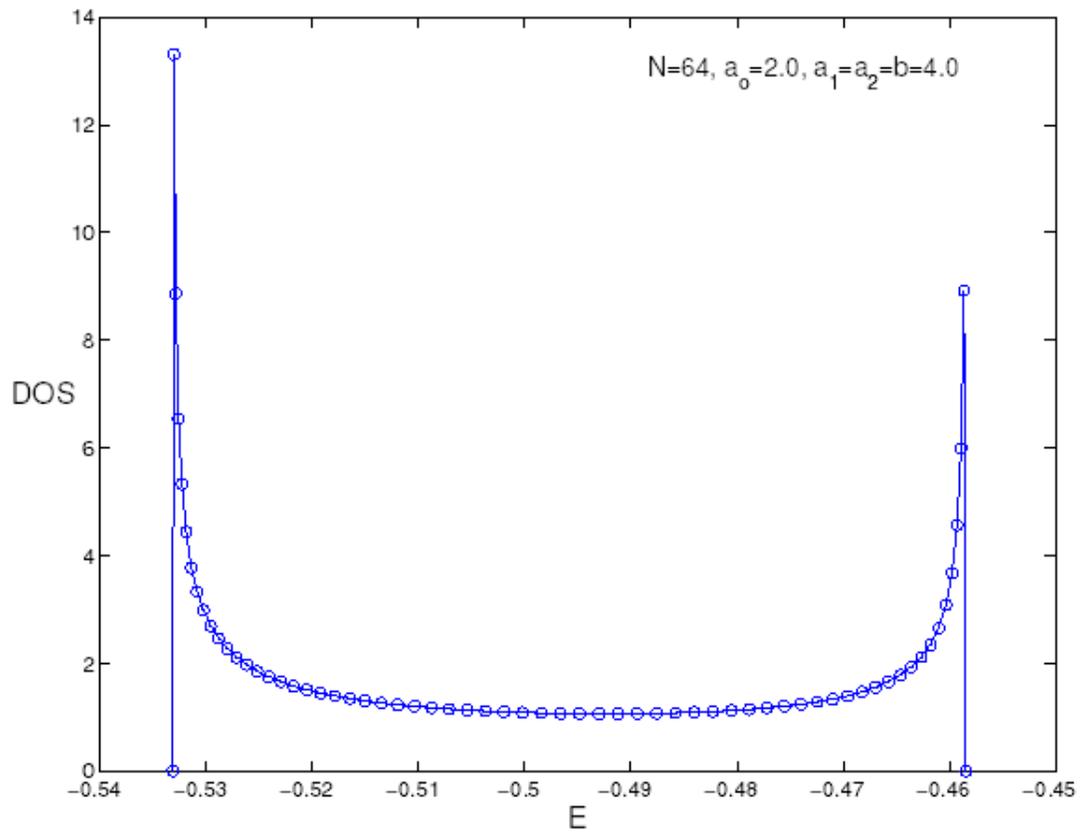

Figure 5a

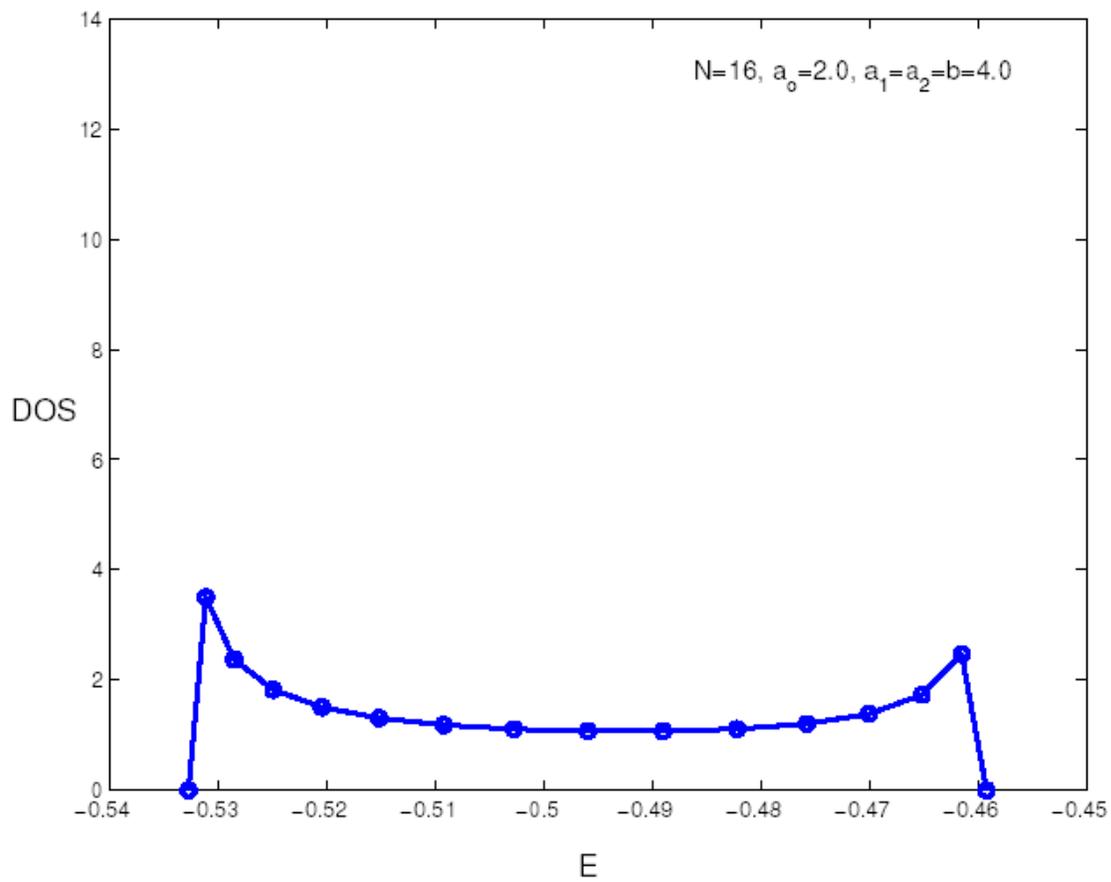

Figure 5b

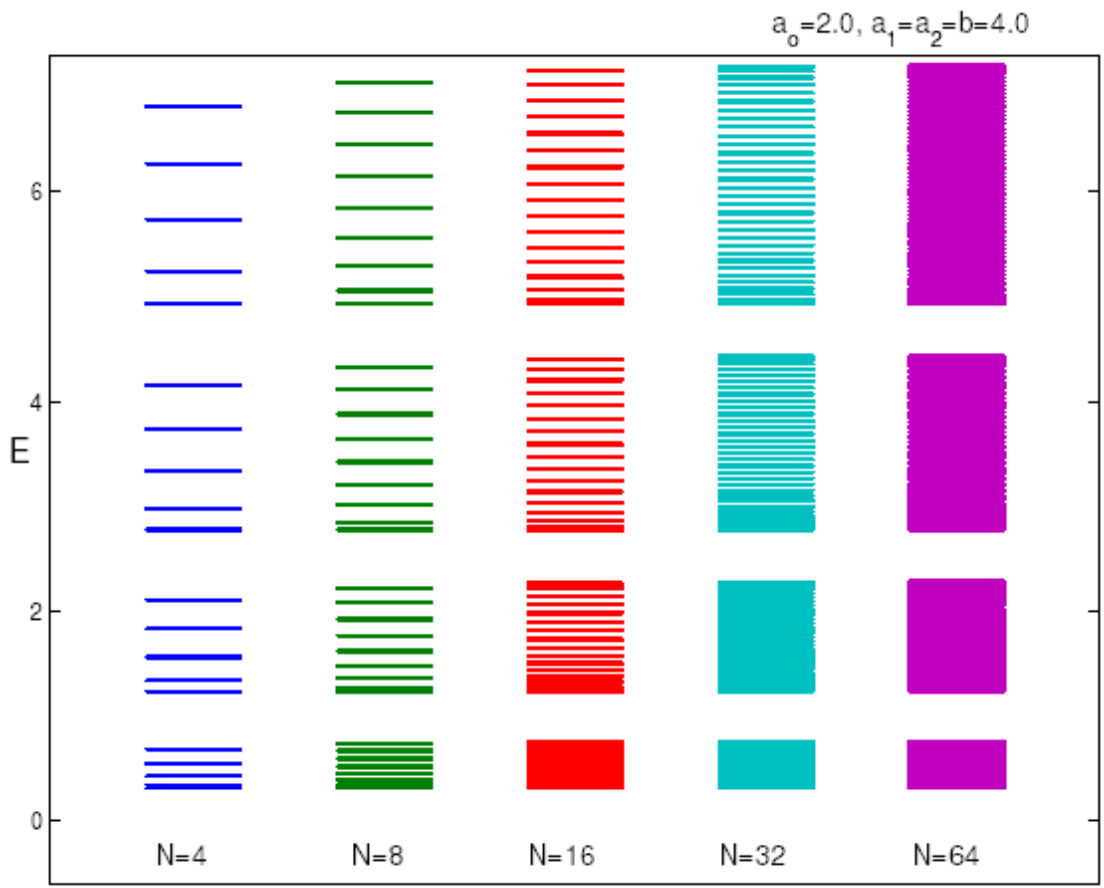

Figure 6

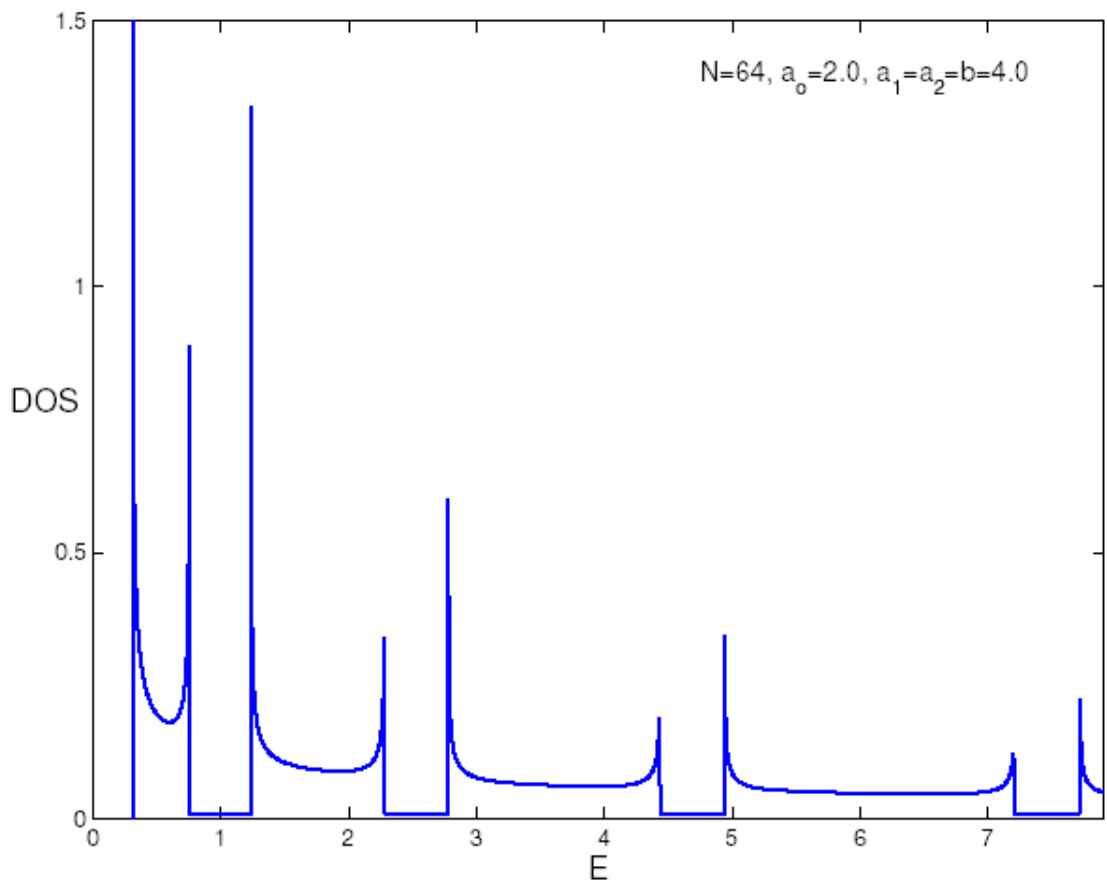

Figure 7

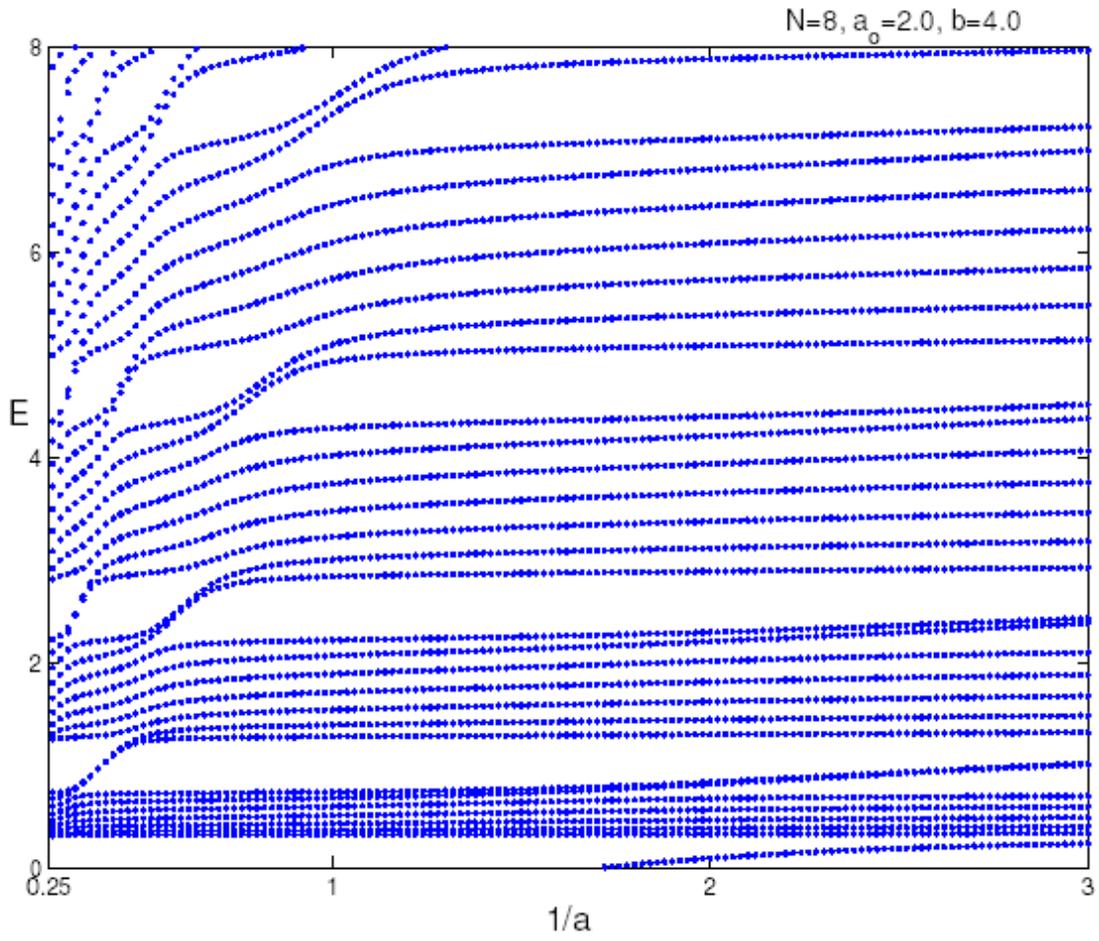

Figure 8a

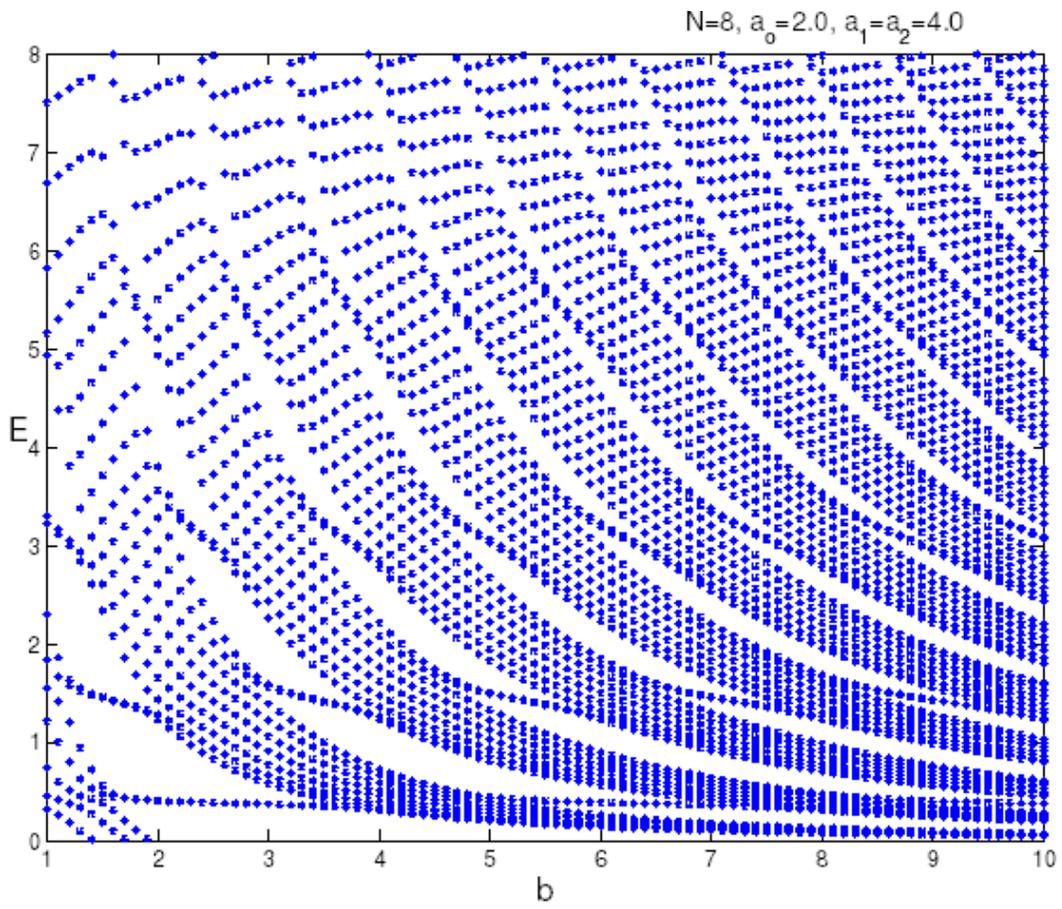

Figure 8b

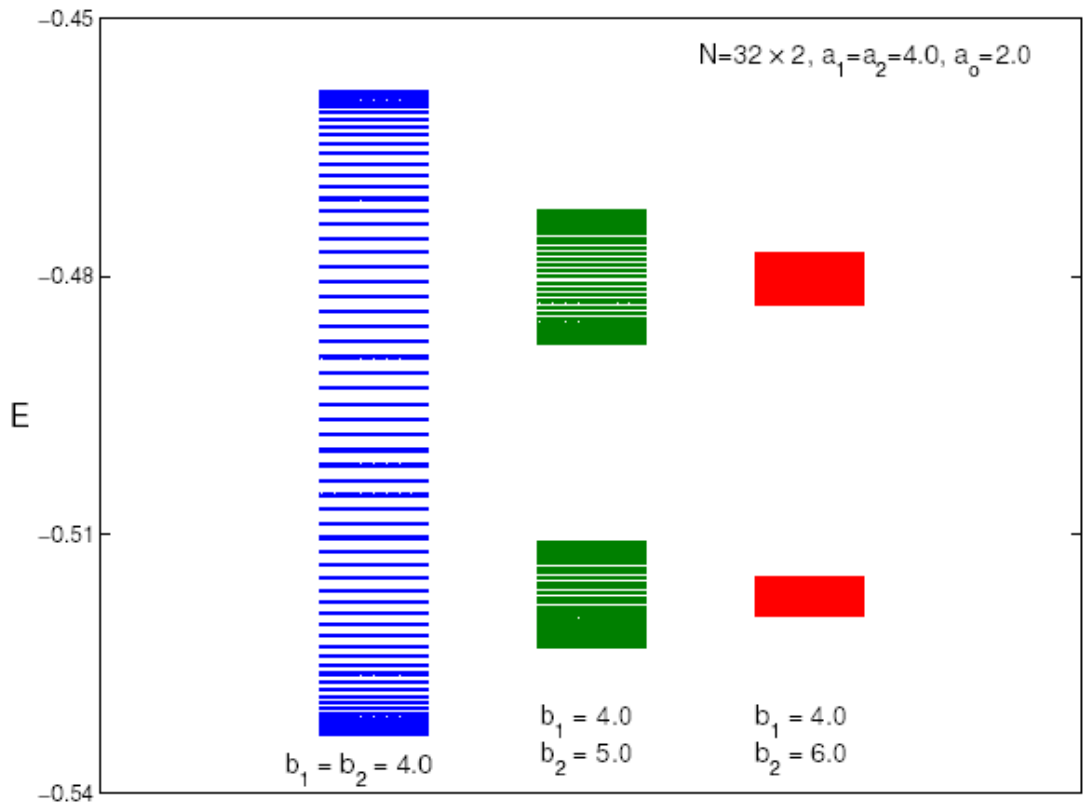

Figure 9a

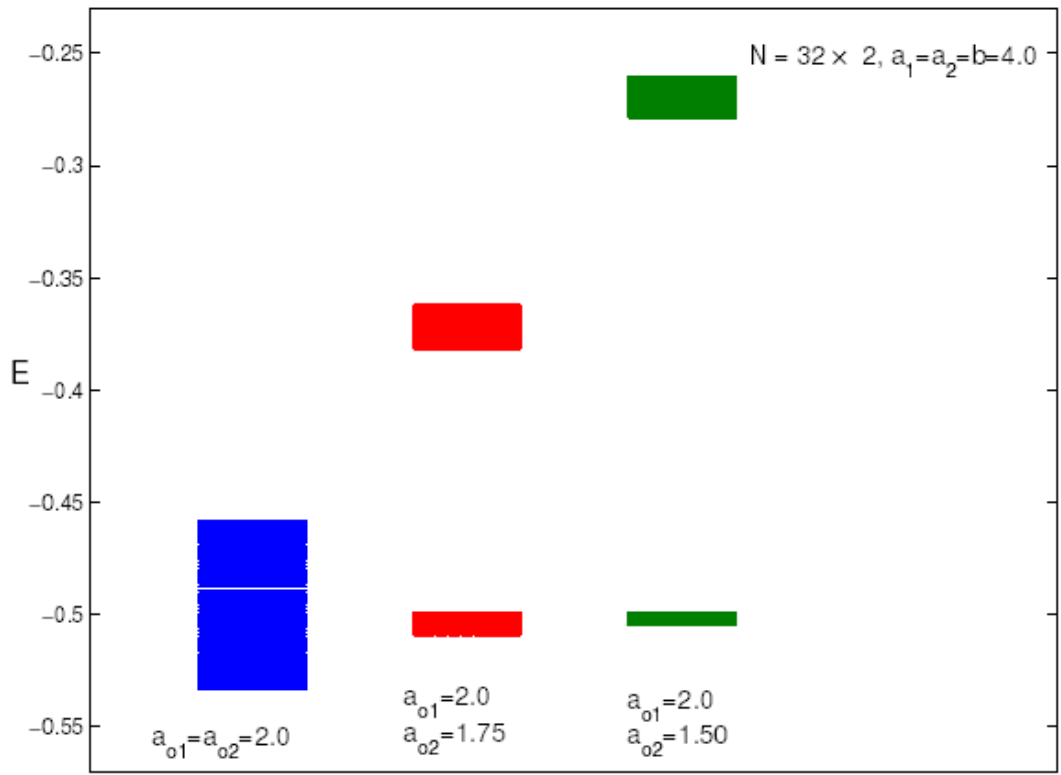

Figure 9b

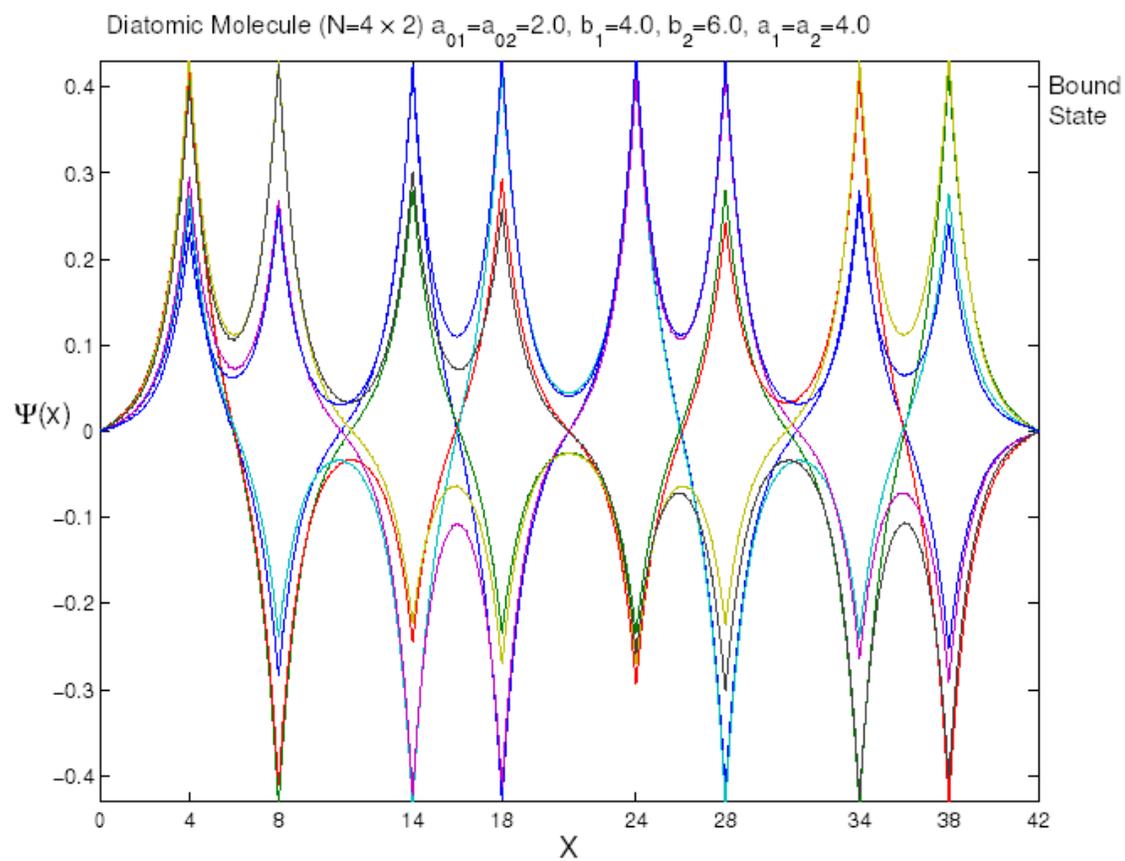

Figure 9c

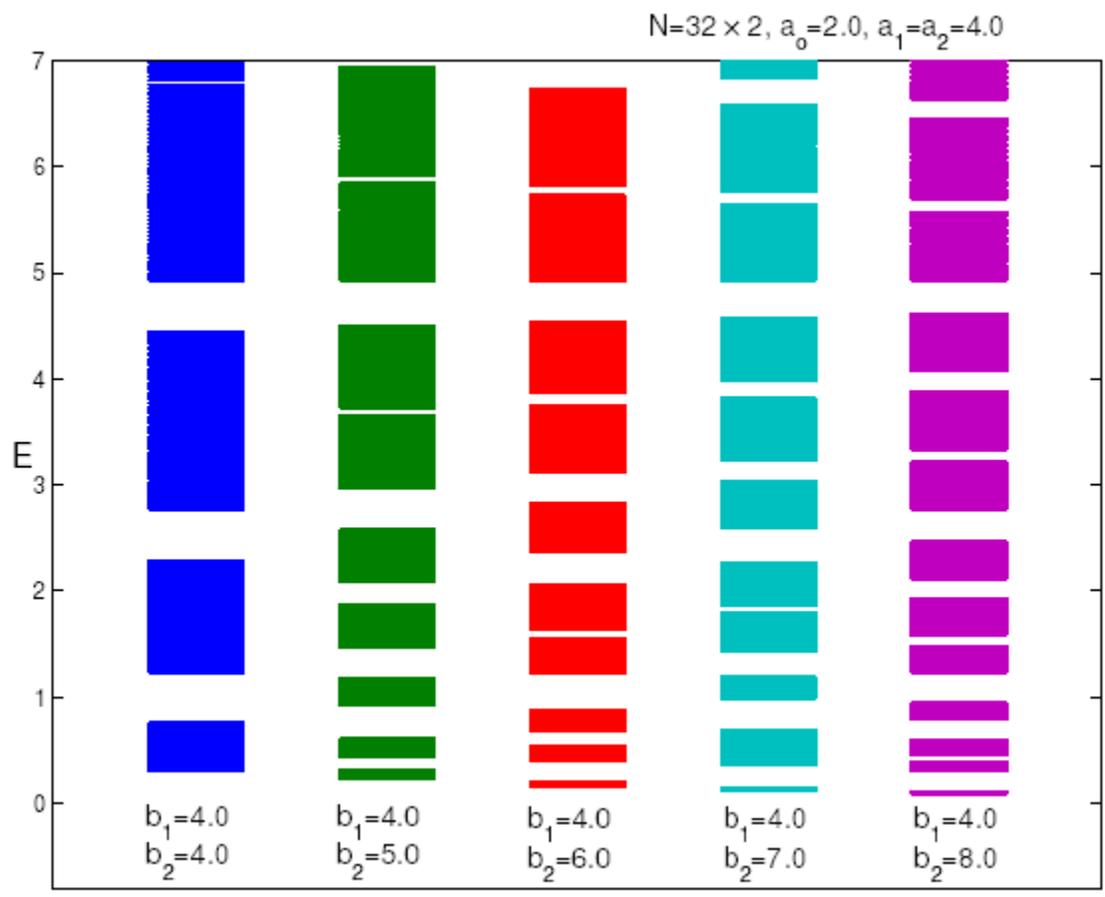

Figure 10

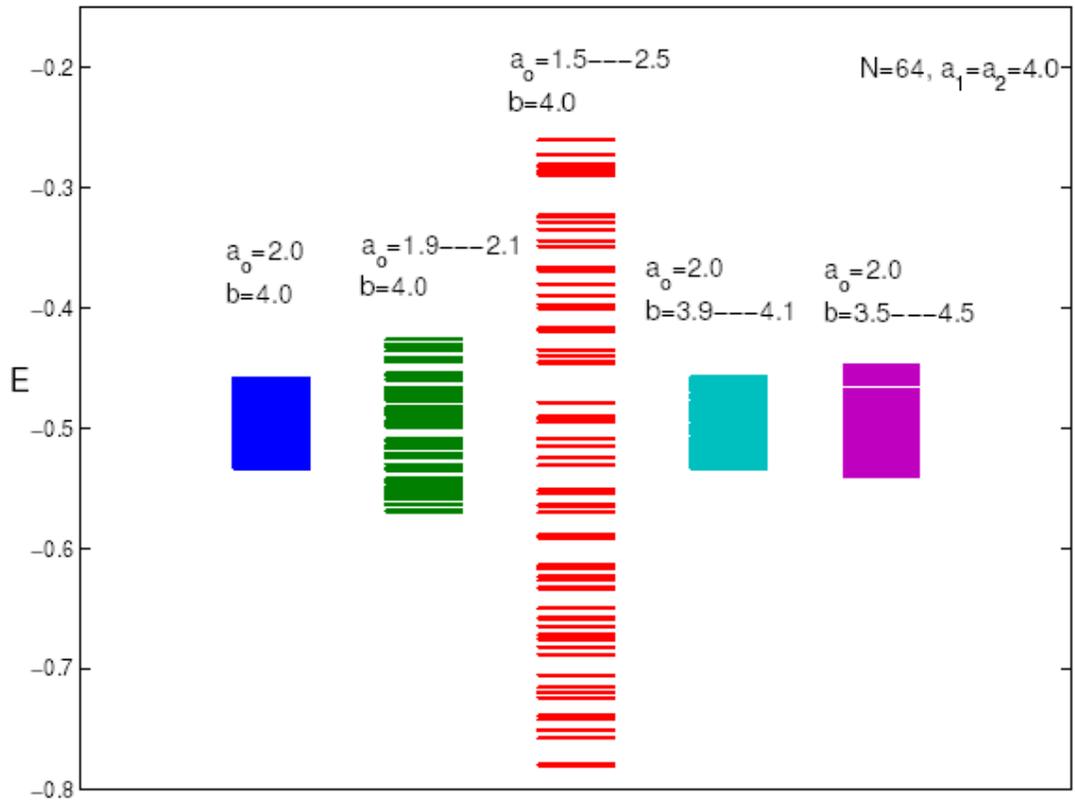

Figure 11

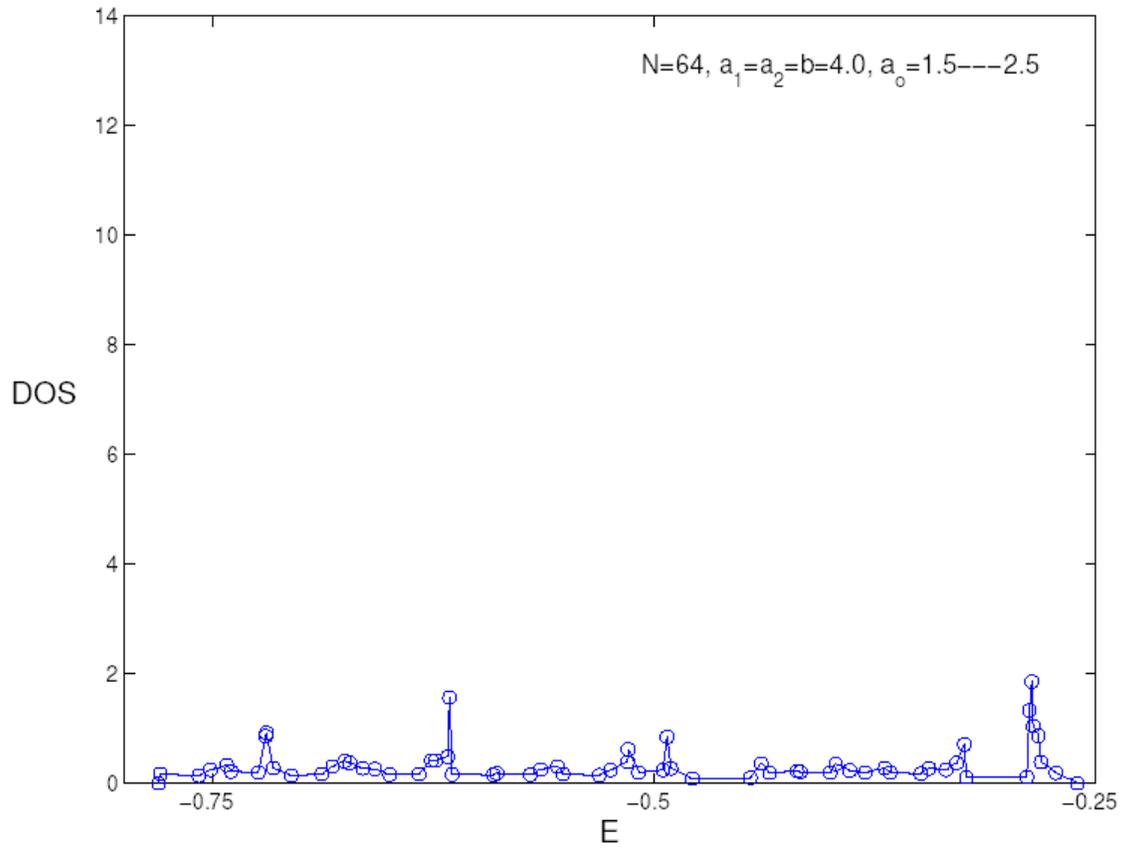

Figure 12a

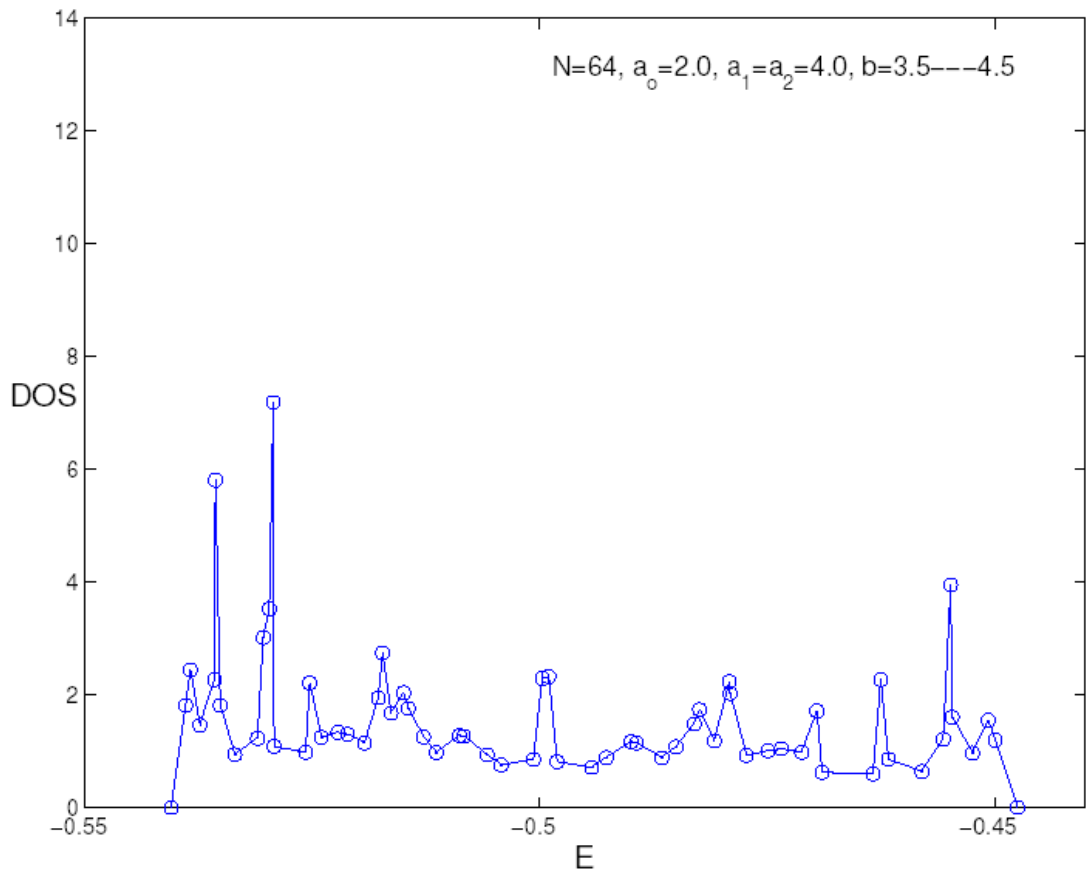

Figure 12b

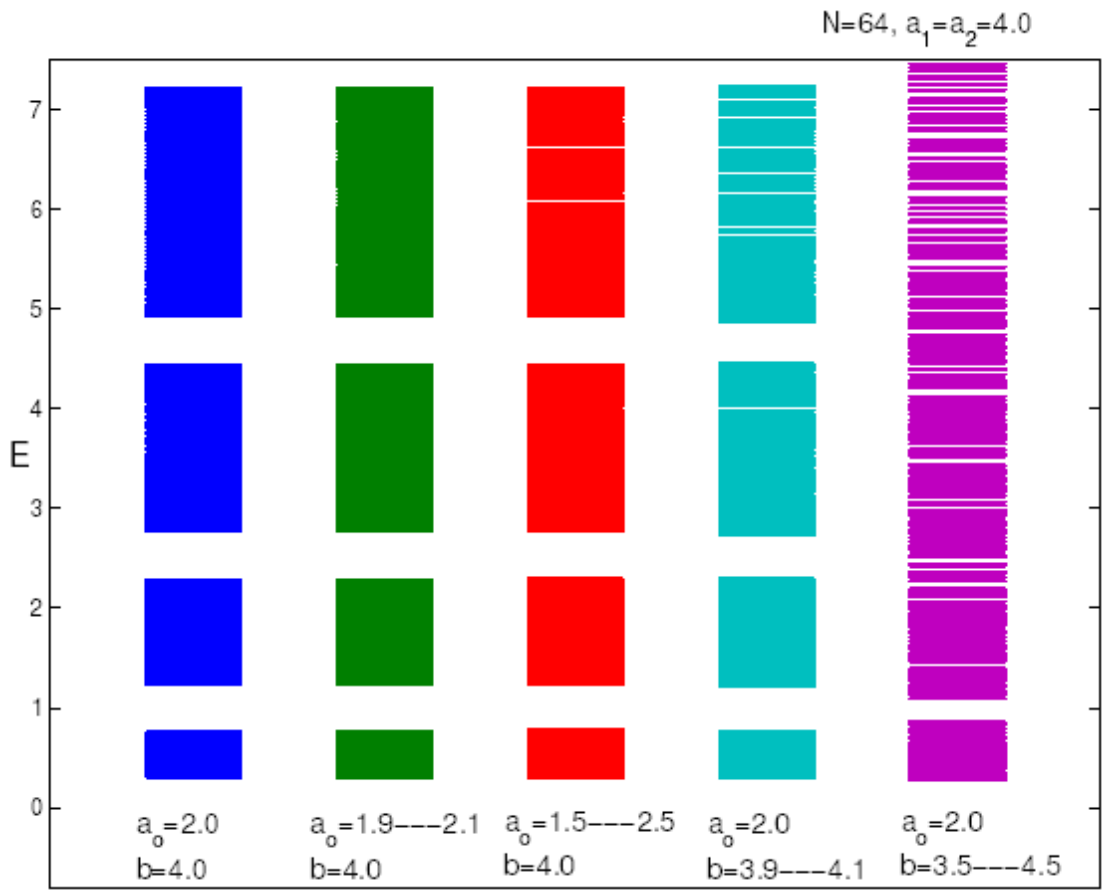

Figure 13

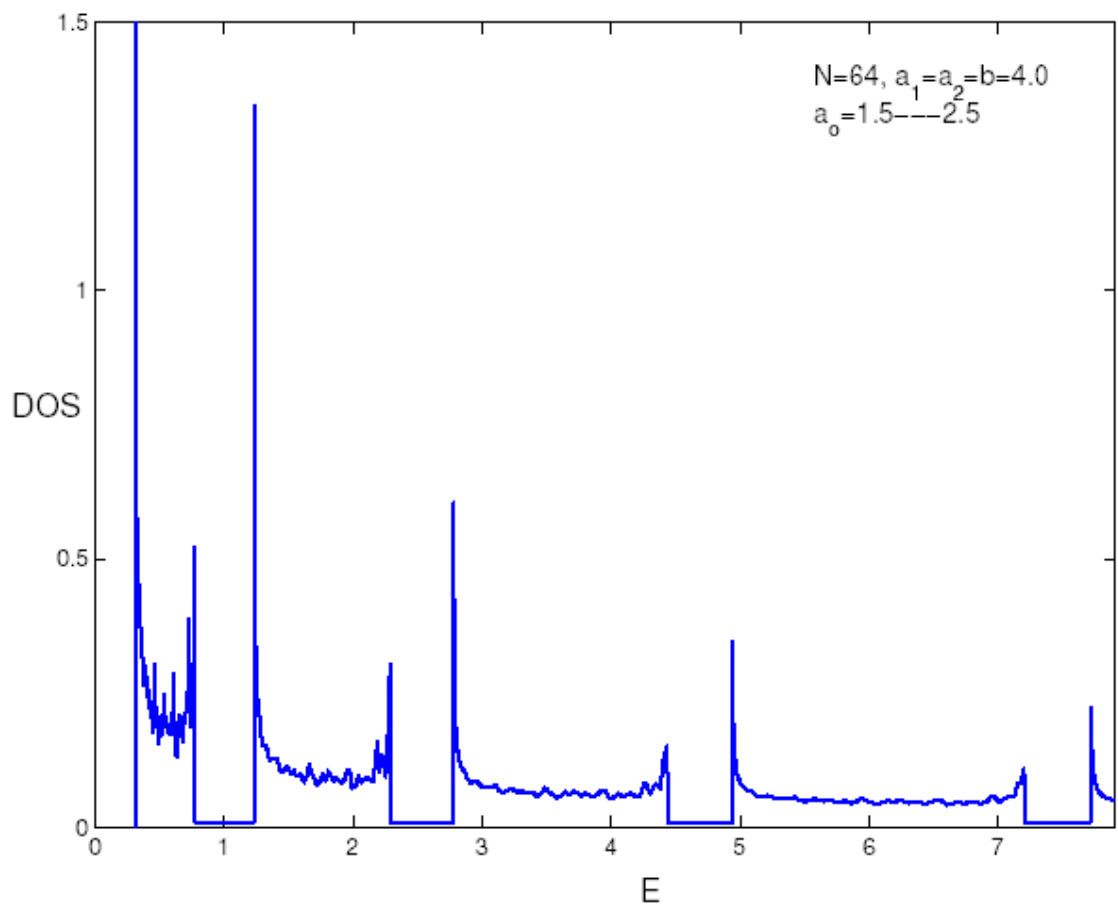

Figure 14a

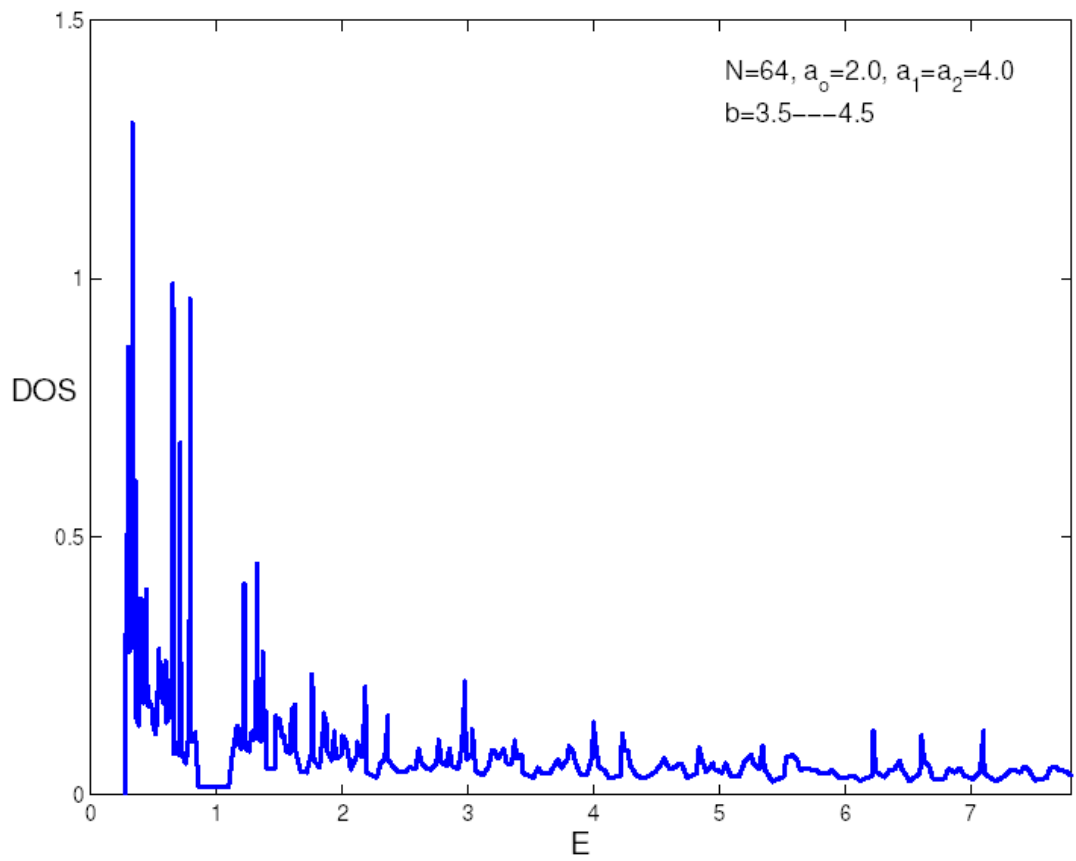

Figure 14b